\documentclass[reprint,
superscriptaddress,
longbibliography,
 amsmath,amssymb,
 aps,
pra,
floatfix,
]{revtex4-2}
\usepackage[utf8]{inputenc}
\usepackage{graphicx}
\usepackage{dcolumn}
\usepackage{soul}
\usepackage{bm}
\usepackage{hyperref}
\usepackage{bm}
\usepackage{physics}
\usepackage{layouts}
\usepackage[english]{babel}
\usepackage{siunitx}
\usepackage[dvipsnames]{xcolor}
\AtBeginDocument{\RenewCommandCopy\qty\SI}
\newcommand{\resp}[2]{#2}

\begin{document}


\title{Phase \resp{D}{d}iagram and dynamical phases of self organization of a Bose--Einstein condensate in a transversely pumped red-detuned cavity}

\author{Julian Mayr}
\affiliation{SUPA, School of Physics and Astronomy, University of St. Andrews, St. Andrews KY16 9SS, United Kingdom}
\affiliation{Physikalisches Institut, University of Bonn, Nussallee 12, 53115 Bonn, Germany}
\author{Maria Laura Staffini}%
\affiliation{SUPA, School of Physics and Astronomy, University of St. Andrews, St. Andrews KY16 9SS, United Kingdom}

\author{Simon B. Jäger}
\affiliation{Physikalisches Institut, University of Bonn, Nussallee 12, 53115 Bonn, Germany}

\author{Corinna Kollath}
\affiliation{Physikalisches Institut, University of Bonn, Nussallee 12, 53115 Bonn, Germany}
\author{Jonathan Keeling}
\affiliation{SUPA, School of Physics and Astronomy, University of St. Andrews, St. Andrews KY16 9SS, United Kingdom}

\begin{abstract}
We study a transversely pumped atomic Bose--Einstein Condensate coupled to a single-mode optical cavity, where effective atom--atom interactions are mediated by pump and cavity photons.  A number of experiments and theoretical works have shown the formation of a superradiant state in this setup, where interference of pump and cavity light leads to an optical lattice in which atoms self-consistently organize.
This self-organization has been extensively studied using the approximate Dicke model (truncating to two momentum states), as well as through numerical Gross--Pitaevskii simulations in one and two dimensions.
Here, we perform a full mean-field analysis of the system, including all relevant atomic momentum states and the cavity field. 
We map out the steady-state phase diagram vs pump strength and cavity detuning, and provide an in-depth understanding of the instabilities that are linked to the emergence of spatio-temporal patterns. 
We find and describe parameter regimes where mean-field predicts bistability, regimes where the dynamics form chaotic trajectories, instabilities caused by resonances between normal mode excitations, and states with atomic dynamics but vanishing cavity field.
\end{abstract}
\maketitle
\section{\label{sec:level1}Introduction}
Atoms coupled to optical cavities have emerged as a versatile platform within which one can explore open-quantum-system physics~\cite{ritsch_cold_2013,mivehvar_cavity_2021}. 
\resp{Theoretical works considering single-mode optical cavities have proposed a variety of steady-state phases as well as dynamical features~\cite{p_domokos_collective_2002,asboth_self-organization_2005,dimer_proposed_2007,keeling_collective_2010,nagy_critical_2011,liu_light-shift-induced_2011,nagy_critical_2011,bhaseen_dynamics_2012,oztop_collective_2013,piazza_boseeinstein_2013,chen_superradiance_2014,keeling_fermionic_2014,piazza_umklapp_2014,pan_topological_2015,piazza_self-ordered_2015,mivehvar_superradiant_2017,molignini_crystallization_2022,gao_self-organized_2023,tuquero_impact_2024,harmon_dynamical_2025}.}{Theoretical works considering single-mode optical cavities have explored how one may realize a generalized Dicke model by Raman driving~\cite{dimer_proposed_2007}, as well as proposing a variety of steady-state phases including spatial atomic self-organization of thermal atoms~\cite{p_domokos_collective_2002,asboth_self-organization_2005}, Bose--Einstein condensates~\cite{keeling_collective_2010,nagy_critical_2011,liu_light-shift-induced_2011,bhaseen_dynamics_2012,piazza_boseeinstein_2013,oztop_collective_2013}, and degenerate fermions~\cite{keeling_fermionic_2014,chen_superradiance_2014,piazza_umklapp_2014,pan_topological_2015,mivehvar_superradiant_2017,molignini_crystallization_2022}.  
Recent areas of interest have included dynamical phases such as limit cycles and chaos~\cite{piazza_self-ordered_2015,gao_self-organized_2023,tuquero_impact_2024,harmon_dynamical_2025}.} 
\resp{Many of these phenomena have been realized and observed experimentally~\cite{black_observation_2003,baumann_dicke_2010,baumann_exploring_2011,mottl_roton-type_2012,brennecke_real-time_2013,klinder_observation_2015,klinder_dynamical_2015,zhiqiang_nonequilibrium_2017,zupancic_p_2019, roux_strongly_2020,zhang_observation_2021,dreon_self-oscillating_2022,kongkhambut_observation_2022,helson_density-wave_2023,skulte_realizing_2024,natale_synchronization_2025,zwettler_cavity-mediated_2025}.}{Many of these phenomena have been realized and observed experimentally.  This includes self-organization with thermal atoms \cite{black_observation_2003,Arnold:2012} and Bose--Einstein condensates ~\cite{baumann_dicke_2010,baumann_exploring_2011}, along with studies of the associated mode softening~\cite{mottl_roton-type_2012} and open  system dynamics~\cite{brennecke_real-time_2013,klinder_dynamical_2015,klinder_observation_2015}. 
Out-of-equilibrium dynamics has been studied in blue-detuned experiments with bosons where the driving leads to repulsive interactions~\cite{zupancic_p_2019,dreon_self-oscillating_2022,natale_synchronization_2025}, as well as with ultra-cold fermions~\cite{roux_strongly_2020,zhang_observation_2021,helson_density-wave_2023,zwettler_cavity-mediated_2025}.  There has been work on the emergence of time-crystals~\cite{Kessler:2021,kongkhambut_observation_2022,skulte_realizing_2024} and including internal spin degrees of freedom~\cite{zhiqiang_nonequilibrium_2017,Kroeze:2018}. 
}
\resp{Further}{Beyond single-mode cavities, further}
 possibilities have been explored more recently using 
crossed optical cavities with two coupled modes~\cite{leonard_supersolid_2017,gopalakrishnan_intertwined_2017}, and with
multimode optical cavities~\cite{gopalakrishnan_emergent_2009,gopalakrishnan_frustration_2011,kollar_adjustable-length_2015,kollar_supermode-density-wave-polariton_2017,vaidya_tunable-range_2018}.

\begin{figure}
    \centering
    \includegraphics[width=\linewidth]{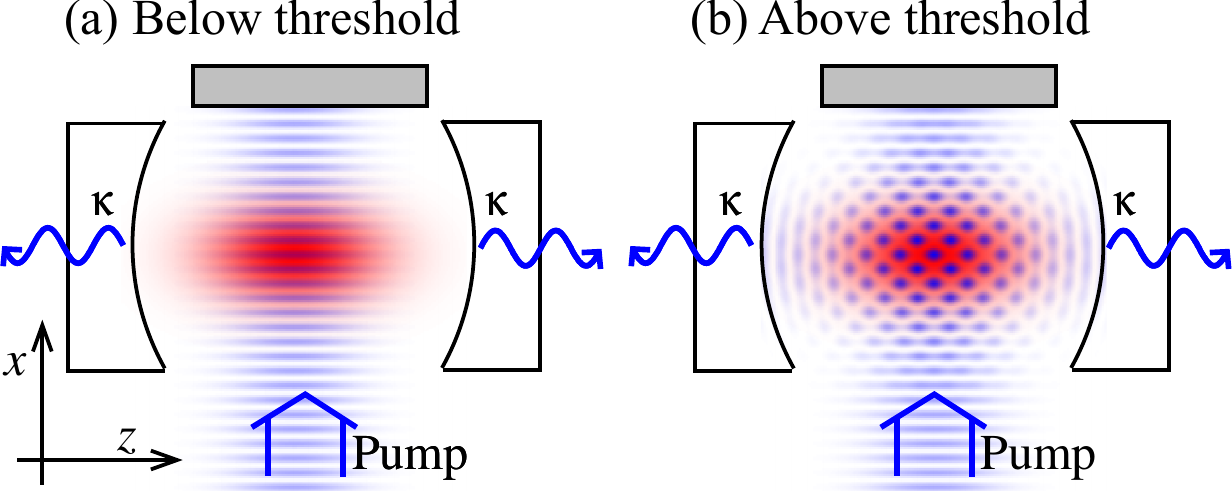}
    \caption{A cartoon of the experiment and the self-organization transition. 
    The atom cloud (a) below and (b) above the critical pumping threshold. Due to the \resp{non-zero}{intracavity} light field \resp{}{which builds up} in the superradiant phase, the atoms organize in a checkerboard pattern. Adapted 
    with permission from Ref.~\cite{bhaseen_dynamics_2012}, Copyright (2012) by the American Physical Society.}
    \label{fig:cavity_transition}
\end{figure}

In the simplest version of such experiments, a Bose--Einstein condensate (BEC) coupled to a transversely pumped single-mode optical cavity undergoes a superradiant phase transition above a critical pumping threshold. In the superradiant phase, the cavity develops a macroscopic photon population due to collective atomic scattering, and the atoms organize in a checkerboard pattern, as illustrated in Fig.~\ref{fig:cavity_transition}. By mapping the system onto a two-state Dicke model~\cite{Hepp1973EquilibriumField}, this can be interpreted as the Hepp--Lieb--Dicke transition to a superradiant state as shown in Ref.~\cite{dimer_proposed_2007}. This was realized experimentally in \resp{Ref.}{2010}~\cite{baumann_dicke_2010}, and \resp{later also in Refs.}{confirmed in further experiments}~\cite{brennecke_real-time_2013, klinder_dynamical_2015}. 
\resp{The dynamics of this system as a Dicke model}{
The mapping of this system to a Dicke model and the resulting dynamics}
were discussed in many theoretical works including Refs.~\cite{keeling_collective_2010,nagy_dicke-model_2010,bhaseen_dynamics_2012,piazza_self-ordered_2015}; for a review, see \resp{}{Ref.}~\cite{Kirton2018IntroductionVersa}.

While the two-state Dicke approximation predicts the phase transition to superradiance, it is known to miss a dynamical phase, resulting from the atom-dependent shift of the cavity resonance.
Such behavior is already visible in the experiment \resp{of}{discussed in} Ref.~\cite{baumann_dicke_2010}: near the bare cavity resonance the cavity photon number in the ``phase diagram'' shows significant variation with detuning, and the time evolution of cavity photon number shows fast variation. The instability is also found numerically in Ref.~\cite{kristian_experimental_2011}. 
For blue detuned cavities, limit cycles and chaos were theoretically predicted in Ref.~\cite{piazza_self-ordered_2015}, while for the red detuned regime, limit cycle behavior was first theoretically found in Ref.~\cite{gao_self-organized_2023}. Both of these works use an approximation neglecting atomic motion along the pump axis, which results in an effectively one-dimensional model. Both regimes are elaborated upon in Ref.~\cite{tuquero_impact_2024} and the impact of quantum noise is analyzed. In Ref.~\cite{skulte_realizing_2024}, limit cycle behavior in the red detuned regime is experimentally realized. Limit cycles are also found in Ref.~\cite{harmon_dynamical_2025}, which demonstrates that adiabatic elimination of the cavity is insufficient to describe their formation. 
In addition to missing the dynamical phases, the Dicke model approximation predicts the phase boundary to the superradiant phase to always be second order.  However it is expected that the superradiant phase transition should become first order in some regime, as reported for fermions in Ref.~\cite{keeling_fermionic_2014}.

The aim of this paper is to systematically map the phase diagram of the red-detuned BEC-cavity system using experimentally relevant parameters, and to understand its behavior in the different phases. To this end, we use a mean field description of the full system, including all relevant momentum states in both pump and cavity direction and the cavity field. To increase our physical understanding of the phases, we use mixed real-time dynamics and steady-state analysis.

\resp{}{
Our results here show that the explicit inclusion of atomic dynamics in the direction of the pump---which is often disregarded---results in experimentally relevant corrections to the phase diagram, e.g. enhancing bistability. 
It also changes the qualitative behavior by inducing  new instabilities toward limit cycle phases that occur via non-Hermitian ``polaritonic resonances''.
Our static analysis of stable and unstable fixed points uncovers bistable phases, which do appear in effective one-dimensional models but have so far been overlooked in dynamical treatments.
Finally we discuss the nature of phases which we refer to as ``stable atomic superpositions'', which can be seen as extensions of the normal phase, where the atoms populate a superposition of single particle eigenstates rather than a single eigenstate.
}

The rest of the paper is organized as follows.  In Sec.~\ref{sec:model_methods} we first introduce the model Hamiltonian, then  derive mean-field equations of motion and describe how we construct the phase diagram using linear stability analysis. In Sec.~\ref{sec:phase_diag}, we then classify the phases and discuss the main features of the phase diagram. In Sec.~\ref{sec:DWres}, we explain the anomalous instabilities of the system resulting from density-wave resonances. We perform dynamical simulations in the unstable region in Sec.~\ref{sec:UnstableDyn}. We find evolution to oscillatory atomic superpositions at zero cavity field, which we explain in Sec.~\ref{sec:SAS}. Section~\ref{sec:conclusion} provides conclusions.

\section{Model and methods}
\label{sec:model_methods}
\subsection{Effective Model}
\begin{figure}
    \centering
    \includegraphics[width=\linewidth]{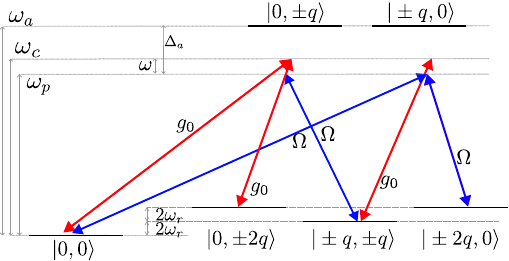}
    \caption{Energy level scheme, showing the possible photon-mediated transitions starting from the $|0,0\rangle$ state. Blue lines denote interaction with a pump photon, red lines interaction with a cavity photon. \resp{}{The energy difference between ground and excited atomic states is $\omega_a$.
    The pump frequency $\omega_p$  is red detuned from this energy by $\Delta_a$. 
    The detuning of the pump from the cavity frequency, $\omega_c$, is denoted by $\omega$. 
    The atomic momentum states are multiples of the cavity recoil momentum $q$ and their energies are thus set by the  corresponding recoil energy $\omega_r={q^2}/{2m}$}.}
    \label{fig:photon_processes}
\end{figure}
We consider the momentum space Hamiltonian of atoms coupled to a single-mode cavity
\begin{multline}
    \hat{H}= \omega \hat{a}^{\dagger} \hat{a}+\sum_{\bm k}\biggl[\frac{k^2}{2 m}  \hat{c}_{\mathbf{k}}^{\dagger} \hat{c}_{\mathbf{k}}
    \\
    -\frac{\Omega^2}{4 \Delta_a} \sum_{s\in\pm 1} \hat{c}_{\mathbf{k}}^{\dagger} \hat{c}_{\mathbf{k}+2 s \mathbf{q}_x}
    -\frac{g_0^2}{4 \Delta_a} \hat{a}^{\dagger} \hat{a} \sum_{s \in\pm 1} \hat{c}_{\mathbf{k}}^{\dagger} \hat{c}_{\mathbf{k}+2 s \mathbf{q}_z} 
    \\
    -\frac{g_0 \Omega}{4 \Delta_a}\left(\hat{a}^{\dagger}+\hat{a}\right) \sum_{s, s^{\prime}\in\pm 1} \hat{c}_{\mathbf{k}}^{\dagger} \hat{c}_{\mathbf{k}+s \mathbf{q}_x+s^{\prime} \mathbf{q}_z}
    \biggr].
\end{multline}
Here $c_{\mathbf{k}}^\dagger$ and $\hat{c}_{\mathbf{k}}$ are the atomic creation and annihilation operators for a boson with momentum ${\bf k}$, and $\hat{a}^\dagger,\hat{a}$ are those for cavity photons. The momentum $\mathbf{q}_x$ and $\mathbf{q}_z$ are the recoil momentum in the pump and cavity directions respectively. The atomic interaction strength with the cavity photons is $g_0$, while $\Omega$ is the Rabi frequency of the driving laser. 
As shown in Fig.~\ref{fig:photon_processes}, the detuning  $\Delta_a$ is that of the pump (at $\omega_p)$ from the internal atomic transition (at $\omega_a)$; this quantity enters the Hamiltonian through adiabatic elimination of excited atomic states. 
When $\Delta_a$ is positive, this describes red detuning, for which the light-induced potential is attractive, hence the signs  of the light-matter interaction in the Hamiltonian.
The effective cavity frequency $\omega$ is the cavity--pump detuning; we follow the sign convention where positive $\omega=\omega_c-\omega_p$ corresponds to negative pump detuning, meaning our phase diagram is flipped in comparison to Ref.~\cite{baumann_dicke_2010}. $\omega_c$ here corresponds to the effective cavity frequency modified by a Stark shift due to the presence of atoms, as described below. 

The three interaction terms correspond to the three possible state-changing two-photon processes.  Figure~\ref{fig:photon_processes} illustrates the states that can be reached via these processes if starting from atoms at rest, i.e.\ the  $|k_x=0,k_z=0\rangle$ state.  As discussed below, our treatment is not restricted to these states; we also include higher motional states which can be reached by repeated application of these three processes. 

The processes can be understood as follows: Absorbing and emitting a pump photon adds two recoil-momentum kicks to an atom, corresponding to the term $V_\text{pump}= \frac{\Omega^2}{4 \Delta_a} \sum_{s\in\pm 1} \hat{c}_{\mathbf{k}}^{\dagger} \hat{c}_{\mathbf{k}+2 s \mathbf{q}_x}$ in the case that absorption and emission momentum gains are aligned. If they are opposite, the state of the atom does not change, and this process corresponds to a constant energy shift that does not contribute to the dynamics. The same picture holds for the interaction with two cavity photons $V_\text{cavity}=\frac{g_0^2}{4 \Delta_a} \hat{a}^{\dagger} \hat{a} \sum_{s\in\pm 1} \hat{c}_{\mathbf{k}}^{\dagger} \hat{c}_{\mathbf{k}+2 s \mathbf{q}_z}$. In contrast to the pump, the cavity photon creation and annihilation is explicitly accounted for. The state-preserving process corresponds to the stark shift absorbed into $\omega_c$. Lastly, the interference term $V_\text{cross}=\frac{g_0 \Omega}{4 \Delta_a}\left(\hat{a}^{\dagger}+\hat{a}\right) \sum_{s, s^{\prime} \in\pm 1} \hat{c}_{\mathbf{k}}^{\dagger} \hat{c}_{\mathbf{k}+s \mathbf{q}_x+s^{\prime} \mathbf{q}_z}$ describes the scattering of pump photons into the cavity and vice versa.

We do not include any contact interaction between the atoms, which we assume to be weaker than the photon mediated interactions. This means that this model should be valid only on short enough timescales, where effects arising due to collisions are negligible. 
At later times, such collisions would be expected to cause thermalisation, redistributing atoms to a range of momentum states rather than just the discrete points in our model.  
In addition, such interactions can cause dephasing, of any state involving a superposition of different atomic states (as occurs for some of the states we discuss below).
The timescale at which those effects become relevant would depend on the density of the atomic gas and the $s$-wave scattering length of the atoms (and thus which atomic species is used).  
In experiments performed to date, models of the form we study appear to match experiments well over the full duration of the experiment, typically a few hundred milliseconds.

In an interaction-free BEC at temperature $T=0$, all atoms are in the $\ket{\psi}=|0,0\rangle$ state.  We assume that the initial state in an experiment is close to such a state.  As such, with the allowed scattering processes, this restricts the system to the subspace of states where $\ket{\psi}=|nq,mq\rangle$ (with $q$ the recoil momentum). We can thus write
\begin{multline}
    \hat{H}= \omega \hat{a}^{\dagger} \hat{a}+\sum_{n,m}\biggl[\omega_r(n^2+m^2)\hat{c}_{n,m}^{\dagger} \hat{c}_{n,m}
    \\
    -\frac{\Omega^2}{4 \Delta_a} \sum_{s\in\pm 1} \hat{c}_{nm}^{\dagger} \hat{c}_{n+2s,m}
    -\frac{g_0^2}{4 \Delta_a} \hat{a}^{\dagger} \hat{a} \sum_{s\in\pm 1} \hat{c}_{n,m}^{\dagger} \hat{c}_{n, m+2s} 
    \\
    -\frac{g_0 \Omega}{4 \Delta_a}\left(\hat{a}^{\dagger}+\hat{a}\right) \sum_{s, s^{\prime} \in\pm 1} \hat{c}_{n,m}^{\dagger} \hat{c}_{n+s,m+s'}\biggr],
\end{multline}
where $\hat c_{n,m}=\hat c_{n\mathbf{q}_z+m\mathbf{q}_x}$ and recoil frequency $\omega_r=\frac{\hbar^2 q^2}{2m}$. The system is additionally confined to states where $n+m$ is even. Restricting the sum to $n_\text{max}=m_\text{max}=1$ results in a Dicke-like Hamiltonian, while restricting only $n_\text{max}=1$, $m_\text{max}\gg1$ results in an idealized one-dimensional model used in some other works~\cite{piazza_self-ordered_2015,gao_self-organized_2023}. In addition to the Hamiltonian dynamics, we assume a cavity loss rate $\kappa$, such that the system obeys the master equation
\begin{equation}
    \frac{d}{dt} \hat{\rho}
    =
    - i[\hat H, \hat \rho]
    + 2\kappa \left(
    \hat a \hat \rho \hat a^\dagger 
    - \frac{1}{2} \{ \hat a^\dagger \hat a, \hat \rho\}
    \right).
    \label{eq:QuantumEOM}
\end{equation}

\subsection{\resp{Mean field equations of motion}{Mean Field Equations of Motion}}
\label{sec:semiclasical}
Since a full solution of Eq.~\eqref{eq:QuantumEOM} is a complex problem, we now treat the time evolution of the system using a mean-field ansatz. For the atoms, we make the Gross--Pitaevskii ansatz, assuming all atoms are condensed into the same single-particle wavefunction,
\begin{eqnarray}
    |\boldsymbol{\phi}\rangle = \frac{1}{\sqrt{N!}}{\left(\sum_{n,m}\phi_{n,m}
    \hat{c}^\dagger_{n,m}\right)}^N |0\rangle,
    \label{eq:GP_state}
\end{eqnarray}
\resp{while f}{with complex coefficients $\phi_{nm}$ defining the macroscopically occupied single-particle mode, normalized as
$\sum_{nm} |\phi_{nm}|^2=1$. F}or the light field, we replace the operator $\hat{a}$ with its coherent state amplitude $\alpha=\langle \hat{a} \rangle$.  This corresponds to making the ansatz $\hat{\rho} = \dyad{\boldsymbol{\phi}} \otimes \dyad{\alpha}$, where $\ket{\alpha}=\exp( \alpha \hat a^\dagger - \alpha^\ast \hat a)\ket{0}$ is a coherent state of the cavity mode. 

This mean-field approach should generally be appropriate for large atom and photon numbers, as it becomes exact in the $N\rightarrow\infty$ limit~\cite{Carollo2021Exactness}. It is known to generally work well for BECs, where typical atom numbers are $N_a=10^5$. The photon number in Ref.~\cite{baumann_dicke_2010} is usually around $N_\lambda=10^2$, which should still be well described by a coherent state, although minor quantum corrections might arise~\cite{muller_genuine_2025}. There are however some regions, especially close to the second-order transition to superradiance, where photon number can be much lower, and a mean-field approach cannot necessarily be trusted. Due to the large cavity linewidth, we expect these regions to be dominated by noise, leading to a smoothing out of second order transitions~\cite{tuquero_impact_2024}.

Inserting this ansatz into the master equation given above we find the coupled differential equations
\begin{multline}
        \frac{d\phi_{n,m}}{dt} =-i \omega_r\left(n^2+m^2\right) \phi_{n, m}
        +
        i \frac{g_0^2|\alpha|^2}{4 \Delta_a} \sum_{s \in \pm 1} \phi_{n, m+2s} \\
         +i \frac{g_0 \Omega\left(\alpha+\alpha^{\ast}\right)}{4 \Delta_a} \sum_{s, s^{\prime}  \in \pm 1} \phi_{n+s, m+s^{\prime}}+i \frac{\Omega^2}{4 \Delta_a} \sum_{s  \in \pm 1} \phi_{n+2s, m}
\end{multline}
\begin{multline}
        \frac{d \alpha}{d t}= -(\kappa+i \omega) \alpha+i \frac{g_0^2 N}{4 \Delta_a} \alpha \sum_{\substack{n,m\\s  \in \pm 1}} \phi_{n, m+2s}^* \phi_{n, m}\\
         +i \frac{g_0 \Omega N}{4 \Delta_a} \sum_{\substack{n,m\\s, s^{\prime} \in \pm 1}} \phi_{n+s, m+s^{\prime}}^* \phi_{n, m}.
\end{multline}
We introduce dimensionless versions of the cavity light field and pump strength
\begin{eqnarray}
    \lambda=\frac{g_0 \alpha}{2\sqrt{\Delta_a\omega_r}},\label{eq:dimless_definitions}
    \qquad
    P=\frac{\Omega}{2\sqrt{\Delta_a \omega_r}}
\end{eqnarray}
and define the Stark shift energy scale
\begin{eqnarray}
    E_0=\frac{g_0^2 N}{4\Delta_a}
\end{eqnarray}
to get
\begin{eqnarray}
    \frac{d \phi_{n, m}}{d t}&=&  -i \omega_r\left[\left(n^2+m^2\right) \phi_{n, m}-|\lambda|^2 \sum_{s \in \pm 1} \phi_{n, m+2 s}\right.\label{eq:EoM_atoms} \\
    &-&\left.P\left(\lambda+\lambda^*\right) \sum_{s, s^{\prime}\in \pm 1} \phi_{n+s, m+s^{\prime}}-P^2 \sum_{s \in \pm 1} \phi_{n+2 s, m} \right],\nonumber \\
    \frac{d \lambda}{d t}&=& -(\kappa+i \omega) \lambda+i E_0 \lambda \sum_{\substack{n,m\\s  \in \pm 1}} \phi_{n, m+2 s}^* \phi_{n, m} \nonumber\\
    & +&i E_0 P \sum_{\substack{n,m\\s, s^{\prime}  \in \pm 1}} \phi_{n+s, m+s^{\prime}}^* \phi_{n, m}.\label{eq:EoM_cavity}
\end{eqnarray}
Practically, the sums over $n$ and $m$ have to be restricted to some maximum momentum to make simulations possible. For the parameters considered in this paper we find that for $n_\text{max}=m_\text{max}=12$, the dynamics are sufficiently converged to make quantitatively accurate predictions (see \resp{App.}{Appendix }~\ref{app:error_analysis} for an analysis of the effect of this truncation), and this truncation will be used in the rest of this paper unless otherwise specified. 

The physical parameter values used in this paper are
$\kappa=2\pi\times \qty{3.98}{MHz}$, $E_0=2\pi \times\qty{6.37}{MHz}$, $\omega_r=2\pi\times\qty{0.0080}{MHz}$. These values approximately match the experiment in Ref.~\cite{baumann_dicke_2010}, but with the  cavity loss rate $\kappa$ increased from $2\pi\times\qty{1.3}{MHz}$ to $2\pi\times\qty{3.98}{MHz}$, making the instabilities discussed in Sec.~\ref{sec:DWres} more pronounced for easier analysis. While we would still observe these instabilities for $\kappa=2\pi\times\qty{1.3}{MHz}$, their decay rates would be significantly smaller, requiring very long simulation time to observe their decay into limit cycles. 

\subsection{Fixed Point Analysis}
\label{sec:LinearStability}

In order to map out the steady state phase diagram, we perform a fixed point analysis \resp{}{\cite{datseris_nonlinear_2022}}.
This fixed-point analysis allows for a clear classification of different phases based on the nature and stability of these fixed points, and is also able to find unstable fixed points, which may contribute to the dynamics. 
This is a much more efficient method than  directly simulating the time evolution described by Eqs.~\eqref{eq:EoM_atoms} and~\eqref{eq:EoM_cavity} for a range of different initial conditions.

Because one of our variables is a wavefunction (specifically the wavefunction of the single-atom state which is macroscopically occupied), we broaden the definition of ``fixed point''  to include a global phase accumulation of that atomic wavefunction.  
As such, a fixed point $(\lambda_0,\bm \phi_0)$ of the equations of motion is defined by
\begin{eqnarray}
    \dot \lambda(\lambda_0,\bm\phi_0) & =&0,\\
    \dot \phi_{nm}(\lambda_0,\bm\phi_0)& =&-i \epsilon_0 \phi_{0,nm},
    \label{eq:fixedpoint_atoms}
\end{eqnarray}
with $\bm \phi_0$ as a vectorized form of the atomic eigenstate $\phi_{0,nm}$ and $\epsilon_0$ being the corresponding (real) eigenenergy.  
Note that the left hand sides of these equations mean the equations of motion as written in Eqs.~\eqref{eq:EoM_atoms} and~\eqref{eq:EoM_cavity}. 
We can exploit the fact that the atoms are not subject to any direct dissipation to rewrite Eq.~\eqref{eq:fixedpoint_atoms} as an eigenvalue equation
\begin{equation}
    \mathbf{H}(\lambda_0)\ket{\bm \phi_0}=\epsilon_0 \ket{\bm\phi_0},
\end{equation}
where $\mathbf{H}(\lambda_0)$ is the atomic Hamiltonian with the constant light field $\lambda_0$ inserted as a parameter.

It is plausible (though not universally guaranteed) to guess that if a stable fixed point exists, the atomic configuration $\bm\phi_0$ will correspond not just to an eigenstate, but specifically to the ground state of the atomic Hamiltonian $\mathbf{H}(\lambda_0)$ for the given cavity field $\lambda_0$. For this part of the discussion, we will focus on this assumption to identify potential steady states. 
This assumption may be checked by direct time evolution of the mean-field equations of motion (\ref{eq:EoM_atoms}) and (\ref{eq:EoM_cavity}), and we find that for most solutions this is true.
However, a regime in which this does not hold will be seen later in Sec.~\ref{sec:SAS}, when we discuss the stability of atomic superpositions which are not ground states (or even eigenstates) of $\mathbf{H}(\lambda_0)$. 

In cases where we can consider the ground state, we require
\begin{equation}
    \bm \phi_0=\text{GS}(\mathbf{H}(\lambda_0))
    \equiv \text{argmin}_{\bm \phi_0} 
    \ev{\mathbf{H}(\lambda_0)}{\bm \phi_0}
\end{equation}
with $\text{GS}(\mathbf{H})$ being defined as the eigenvector of $\mathbf{H}$ with the lowest eigenvalue. We can then find the fixed points by solving the combined equation for $\lambda_0$:
\begin{equation}
    \label{eq:GS_condition}
    \dot \lambda(\lambda_0, \text{GS}(\mathbf{H}(\lambda_0)))=0.
\end{equation}
Equation~\eqref{eq:GS_condition} can be solved numerically by standard root-finding algorithms, \resp{}{we here choose the trust-region algorithm as implemented in the \emph{NLsolve.jl} library.}

To find all possible fixed points, we initialize the root solver at a large number of randomly sampled $\lambda$. We use the trust region method with a tolerance of $\text{Re }\dot\lambda,\text{ Im }\dot \lambda <10^{-11}$. While it might be possible to miss fixed points with a very small basin of attraction in the solver dynamics, we do not find any unexpected additional fixed points in any sweep, even with a large number of initial conditions. This suggests that our search is sufficiently exhaustive that all fixed points are being found.

The differential equation $\dot \lambda=\dot \lambda(\lambda,\text{GS}(\mathbf{H}(\lambda)))$ 
describes pseudodynamics similar to imaginary-time evolution for the atoms and real-time evolution for the cavity field.
\resp{a strategy}{Such a mixed real-time imaginary-time strategy was}
used to find the steady state in Ref.~\cite{harmon_dynamical_2025}. 
\resp{}{Note however that our approach is not identical to this mixed dynamics:}
The dynamics our equation describes is one in which atoms are always constrained to the ground state of the instantaneous potential.  
This is the limiting case of mixed real-time imaginary-time evolution in which the imaginary-time evolution for the atoms is allowed to run to completion (relaxation to the zero temperature state) at each timestep before the next real-time evolution of the cavity field.
As compared to the general mixed real-time imaginary-time evolution approach, our approach takes advantage of the explicit form of the atomic problem as being defined by a single-particle Hamiltonian.
In addition, our approach also finds unstable fixed points, or those \resp{with a small basin of attraction}{to which the dynamics would be unlikely to converge}.

After finding the fixed points, we determine their stability within linear stability analysis. We make the ansatz
\begin{eqnarray}
    \lambda&=&\lambda_0+\delta\lambda,\\
    \bm \phi&=&e^{-i\epsilon_0t}(\bm \phi_0+\delta\bm{\phi}).
\end{eqnarray}
The additional phase factor in $\bm \phi$ removes the ground-state energy-dependent phase rotation from the dynamics. After inserting the ansatz into Eqs.~\eqref{eq:EoM_atoms} and~\eqref{eq:EoM_cavity} and linearizing in the perturbations, we parameterize the complex perturbations as
\begin{eqnarray}
    \delta\lambda&=&xe^{-i\eta t}+y^*e^{i\eta^* t}\label{eq:lambda_to_xy}\\
    \delta\phi_{nm}&=&a_{nm}e^{-i\eta t} + b_{nm}^* e^{i\eta^* t}\label{eq:phi_to_ab},
\end{eqnarray}
with $\eta$ corresponding to the complex frequency, i.e. energy and decay rate, which is found as an eigenvalue of the equations of motion as we discuss next. This parameterization---analogous to that used to derive Bogoliubov--de Gennes equations\resp{}{\cite{gennes_superconductivity_2018}}---is required to treat the complex conjugate variables since the equations of motion couple terms dependent on $\lambda$ and $\lambda^*$. When inserting the ansatz, the equations can be separated according to  their time dependence. We can then write all variables as a vector 
\begin{equation}
    \label{eq:Mevecdef}
    \mathcal{V}=\begin{pmatrix}
    \bm a\\\bm b\\x\\y
    \end{pmatrix},
\end{equation}
and the linearized equations as
\begin{equation} 
\eta\mathcal{V}=\mathcal{M}\mathcal{V}.\label{eq:M_definition}
\end{equation}
This equation defines the allowed values of $\eta$ as the eigenvalues of the matrix $\mathcal{M}$. Calculating the elements of $\mathcal{M}$ is straightforward, and they are given in Appendix~\ref{app:matrix_elements_calc}.

The stability of the fixed points then follows directly from the eigenvalues of $\mathcal{M}$: any eigenvalue with a positive imaginary part implies an exponential growth of the perturbations in Eqs.~\eqref{eq:lambda_to_xy} and~\eqref{eq:phi_to_ab}, and therefore an instability of the fixed point.

\section{\label{sec:phase_diag}Phase diagram}

\begin{figure*}[ht]
    \includegraphics[width=\textwidth]{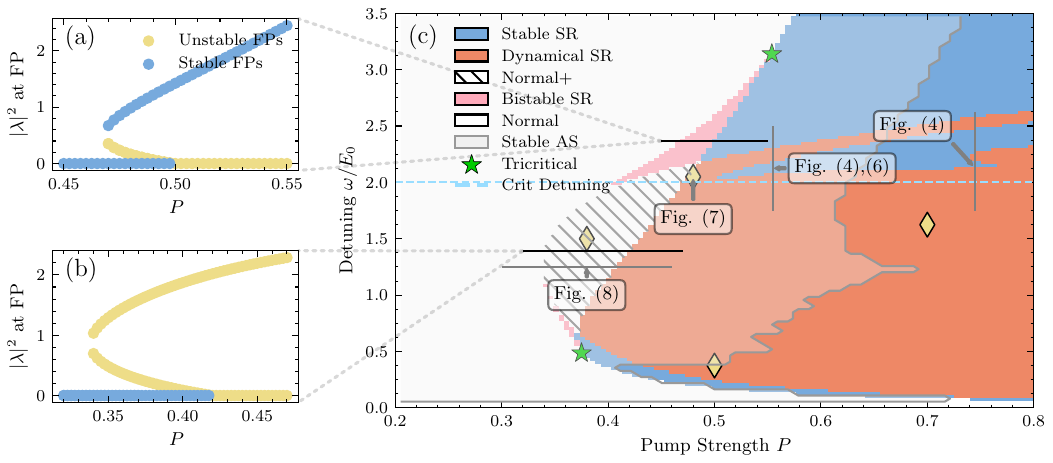}
    \caption{\resp{Right:}{(c)} The phase diagram of the model as a function of dimensionless pump strength $P$ and pump-cavity detuning $\omega$. \resp{Left}{(a)--(b)}: the cavity field amplitude $|\lambda|^2$ of the fixed points and their stability along two cuts marked by solid black lines in the phase diagram. Additional cuts corresponding to subsequent figures are marked as gray lines, and specific points at which dynamics is shown later are marked as yellow diamonds.}\label{fig:full_diagram}
\end{figure*}

To construct the phase diagram, we divide the phase diagram into pixels over a range of detunings $\omega$ and \resp{}{pump strengths} $P$, and find all fixed points for each pair of given parameters. The experimentally fixed cavity decay rate $\kappa$ and stark shift scale $E_0$ are kept constant throughout, at the values given in Sec.~\ref{sec:semiclasical}. We then determine the stability for each fixed point found. This results in different numbers of fixed points, as well as different stability properties of each. We find five qualitatively different configurations, which define five different phases. Each configuration occurs for some value of $P$ in at least one of the left panels of Fig.~\ref{fig:full_diagram}.
\resp{}{We define these phases as follows:}

\begin{description}
    \item[Normal phase] In this phase, there is a stable fixed point at $\lambda=0$ and no other fixed points. Therefore for $\lambda=0$, the cavity is empty and the atomic state corresponds to a BEC with density wave order only in the pump direction.
    \item[Stable superradiant phase] There is one pair of stable fixed points at $\lambda=\pm|\lambda|\neq 0$, and no other stable fixed points  (due to the symmetry $\lambda\rightarrow-\lambda$, fixed points at non-zero $\lambda$ always appear in pairs). Specifically, the fixed point at $\lambda=0$ is unstable. Therefore, when quenched into this phase from the normal phase, the system should become superradiant and converge towards a density-wave ordered superradiant steady state.
    \item[Bistable Superradiant phase] There are stable fixed points both at $\lambda=0$ and at $\lambda\neq 0$, as well as an additional pair of unstable fixed points.  As one moves through this phase, the unstable fixed point(s) connect between the stable fixed points such that there is no true discontinuity of fixed point positions. The state is classically bistable (see later for discussion beyond the classical limit).  As such, which steady state is found depends on initial conditions.
    \item[Dynamical Superradiant phase] There is one fixed point at $\lambda=0$ and one pair at $\lambda\neq 0$, but none of them are stable. Therefore, within the classical approximation, the system in this phase must remain dynamical and can never converge to a steady state.
    \item[Normal+ phase] As in the bistable phase, there are two fixed points pairs at $\lambda|\neq 0$, however now the only stable fixed point is at $\lambda=0$.
    \resp{}{As such, the steady state behavior in this phase is the same as for the normal phase, however the dynamics can be different.}    
    \resp{This}{This phase} can be seen as the overlap of the bistable and dynamical phase, where the superradiant phase has already become unstable but the normal phase remains. \resp{The steady state behavior is the same as for the normal phase.}{} 
\end{description}
The complete phase diagram is shown in Fig.~\ref{fig:full_diagram}. In addition to the phases classified above, the tricritical points, where first order transitions become second order, and the stable atomic superposition (SAS) phase (discussed in Sec.~\ref{sec:SAS}) are marked. The remainder of this section summarizes the key features of this phase diagram, with more detailed discussions of the underlying physics of each different phase provided in subsequent sections.

\paragraph{Bistability.}
Between the stable superradiant (SR) and the normal phase, we find a region of bistability of normal and SR phases. 
The bistable region vanishes, and the transition becomes second-order above the tricritical point marked by a star in Fig.~\ref{fig:full_diagram}. The appearance of the bistable region is a direct consequence of including higher momentum states in the model. 

Beyond mean-field theory, when considering the true steady-state density matrix, this region is expected to be replaced by a sharp first-order transition between the superradiant and normal phase, due to tunneling between the coexisting states. The true phase boundary will therefore be somewhere inside in the bistable region.
For completeness, we briefly summarize here the standard picture~\cite{minganti_spectral_2018,fazio_many-body_2024} of how this occurs.
The full quantum solution for the system density matrix describes the ensemble average over all experiments.
When tunneling between two states can occur, one might anticipate that the full quantum density matrix involves a mixture of those two states, determined by the ratio of the tunneling rates.
That is, if one has states $A,B$ and tunneling rates $r_{A\to B}$ and $r_{B\to A}$, then in a steady state, the probabilities of being in states $A,B$ will obey $p_A/p_B = r_{B \to A}/r_{A \to B}$.

Considering this full quantum solution, as system size becomes large, one finds that two things occur:  First, the tunneling rates become small, so that some sense of bistability still occurs, in that it takes a long time to switch away from the initially prepared state.
Secondly, the ratio of tunneling rates also diverges or vanishes.  That means that the ensemble average becomes dominated by one or other state, thus corresponding to a first-order transition.
This divergence or vanishing of the ratio of tunneling rates is entirely consistent with the vanishing tunneling rate.
One simple model is to consider a case where $r_{A \to B} = r_0 \exp( - N w_{A \to B})$ (and similarly for $B \to A$), where $N$ denotes system size.  
For this form the ratio of tunneling rates will give
$p_A/p_B = \exp[ - N (w_{B \to A} - w_{A \to B})]$.  In this case, if $w_{B \to A} > w_{A \to B}$ then the ratio will go to zero at large $N$, while for $w_{B \to A} < w_{A \to B}$ the ratio will vanish at large $N$.
Exactly this kind of behavior was found numerically in various systems, including for generalized Dicke models~\cite{muller_genuine_2025}.

Therefore, bistability would still be experimentally observable at large atom numbers due to long switching times between the states, with the bistable region still playing a role as defining where these long-lived states exist.
This form of bistability could be consistent with the observation of hysteresis at the superradiance transition in Ref.~\cite{klinder_dynamical_2015}.

\paragraph{Extended Chaotic Region at $\omega<2E_0$.}
The instability of the superradiant phase below $\omega=2E_0$ (where $E_0$ is the Stark shift energy defined previously) is a known feature of atom-cavity systems that goes beyond the simple Dicke model. It results from the atom-ordering-induced shift of the effective cavity frequency~\cite{mivehvar_cavity_2021}:
\begin{equation}
    \omega_\text{eff}=\omega-E_0\sum_{n,m,s} \phi_{n,m+2s}^*\phi_{n,m}.
\end{equation}
For $\omega\leq 2E_0$, self-organization can lead to $\omega_{\text{eff}} \approx 0$. This tunes the pump and cavity to resonance, destabilizing the self-organization. Evidence of this effect has been experimentally observed in Ref.~\cite{baumann_dicke_2010}. The dynamics in this phase appear chaotic. When talking about chaos in this paper, we mean chaos in the sense that two mean-field trajectories at slightly perturbed initial conditions will exponentially diverge from each other. We discuss this more rigorously in Sec.~\ref{sec:UnstableDyn} by introducing Lyapunov exponents. This mean-field behavior has been observed in similar models, like the unbalanced~\cite{stitely_nonlinear_2020} and anisotropic~\cite{mondal_transient_2025} Dicke model. Reference~\cite{mondal_transient_2025} also highlights the corresponding quantum behavior, which we would expect to be similar in this system.

\paragraph{Normal+ phase.}
Due to superradiant fixed points becoming unstable below $\omega<2E_0$, the additional fixed points in the bistable phase also become unstable. This leads to a phase where, as in the normal phase, only the $\lambda=0$ fixed point is stable, but two additional unstable pairs of fixed points appear. While this has no impact on the steady state inside this phase, it leads to the transition to the dynamical phase becoming discontinuous as the chaotic attractors replacing the $\lambda=0$ fixed point are already far away from it. This is in contrast to the transition at lower $\omega$ described below. It also leads to interestingly shaped bistabilities, where when increasing $P$, the superradiant fixed point may become unstable before becoming stable again.

\paragraph{Unstable tongues at \texorpdfstring{$\omega>2E_0$}{large cavity detuning}.}
The stable superradiant regime at $\omega>2E_0$ is interrupted by regions of instability. These are not caused by the standard instability mechanism of atomic ordering shifting the effective cavity frequency, but instead result from resonances of the system's coupled light-matter normal modes. We explain this in detail in Sec.~\ref{sec:DWres}. These instabilities have low growth rates and they are narrow stripes in parameter space, explaining why they have not been observed in Ref.~\cite{baumann_dicke_2010}.

\paragraph{Stable superradiance at \texorpdfstring{$\omega\ll 2E_0$}{low cavity detuning}.}
At low $\omega$, where no normal+ phase exists, a small region of stable superradiance appears. \resp{There, the transition is second order, or the first order jump at the transition is minimal, and the atomic ordering at the fixed points immediately after the transition is too weak to sufficiently modify the effective cavity frequency}{From the numerical results it is not clear whether the transition to this region from the normal phase is second order, or whether it is very weakly first order (i.e. first order with a small jump of the order parameter $\lambda$). 
The smallness of $\lambda$ near this transition is why this superradiant state remains stable, in contrast to other regions at $\omega < 2E_0$:  The small light field in the region close to the transition is insufficient to tune the effective cavity frequency into resonance with the pump, so the superradiant fixed point remains stable}. 
This stable region also extends into the normal+ phase through a second tricritical point to create a small bistable region at low pump strengths.

\paragraph{Stable Atomic Superpositions (SAS).}
As discussed below and in Ref.~\cite{gao_self-organized_2023}, by considering time evolution dynamics,  under some conditions there can be convergence to states that have $\lambda=0$ but contain oscillating atomic superpositions, see Fig.~\ref{fig:SAS_convergence}. The linear stability treatment discussed so far only considers the stability of the ground state of the single-atom problem, so it neglects these states. 
Any superposition of states occupying momentum bins of only even $n,m$ is a solution of the equation of motion consistent with $\lambda=0$, since the cavity couples only to the overlap of adjacent momentum states. In Sec.~\ref{sec:SAS}, we use Floquet analysis to find that there is a continuum where these superposition states are \emph{stable} in large regions of the parameter space, marked as a shaded region in Fig.~\ref{fig:full_diagram}. We call this region of parameter space the stable atomic superposition (SAS) region. These stable superpositions found are bistable with the steady states in their respective regions (if any exist).

\begin{figure*}
    \centering
    \includegraphics[width=\linewidth]{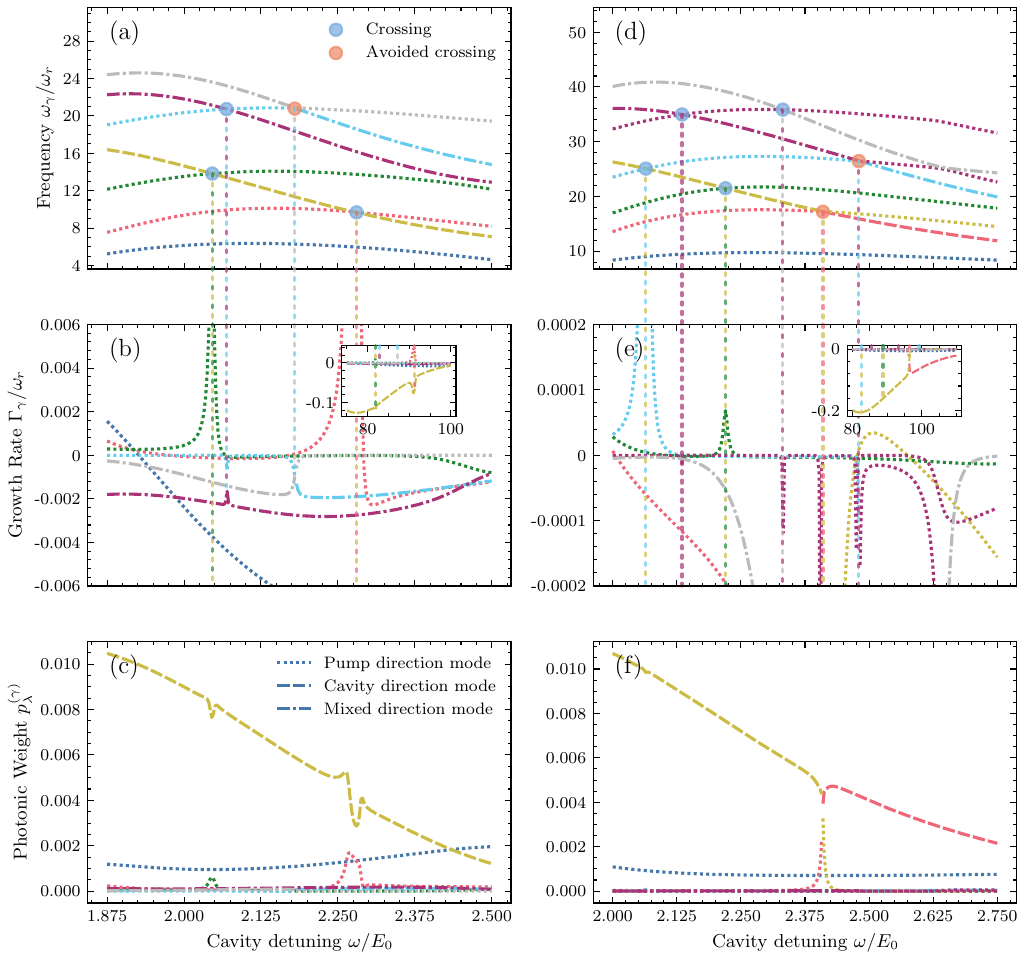}
    \caption{Linear stability mode analysis along the cuts at \resp{(a)}{(a)--(c)} $P=0.55$ and \resp{(b)
    }{(d)--(f)} $P=0.75$ indicated in Fig.~\ref{fig:full_diagram}. \resp{Top:}{(a)/(d)}: Energies $\omega_\gamma$ of the modes while varying $\omega$. \resp{Middle}{(b)/(e)}: The corresponding growth/decay rates ($\Gamma_\gamma$, same color), zoomed out in inset. \resp{Bottom:}{(c)/(f)}: The photonic participation factor (see Eq.~\eqref{eq:participation_factor}) of the modes. At mode crossings, mixing of the eigenvectors can modify the photonic part of otherwise mostly density-wave like modes, changing their decay rates. 
    The modes' directions are identified by comparing their energies to the eigenenergies of 1D cosine potentials, see App.~\ref{app:mode_class}.
    }\label{fig:modes}
\end{figure*}

\begin{figure}[htbp]
    \centering
    \includegraphics[width=\linewidth]{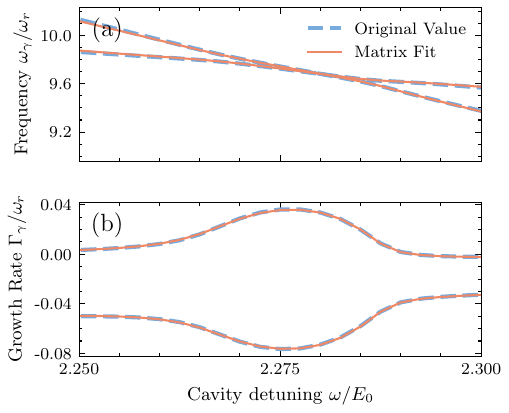}
    \caption{A fit of the matrix eigenvalues of Eq.~\eqref{eq:fitmatrix} to the energies and growth/decay rates of the resonance at $\omega/E_0 \approx 2.25$, $P=0.55$ (see Fig.~\ref{fig:modes}. Fit parameters found are
    $\alpha_1=(7.03+0.0053i)\times 10^{-3}$, $\beta_1=2\pi\times(-25.4 - 0.0094 i)$ MHz,$\alpha_2=(1.96+0.0303i)\times10^{-2}$, $\beta_2=2\pi\times(-18.5 + 0.045 i)$ MHz, $\delta=2\pi\times(0.0466 + 0.423 i)\times10^{-3}$ MHz.}
    \label{fig:matrix_fit}
\end{figure}

\section{Instabilities from Polariton Resonances\label{sec:DWres}}

This section aims to explain the nature of additional instabilities observed in the phase diagram (see Fig.~\ref{fig:full_diagram}) for $\omega > 2E_0$. These features originate from resonances between different collective modes of the coupled atom-cavity system. Due to the coupling between atomic and cavity degrees of freedom, all normal modes of this system are superpositions of atomic and photonic excitations, and thus can be called polaritons. In our parameter regime, the low-energy excitations are of atomic density-wave character, with only a small participation of photonic excitations.

We find that each ``tongue'' of instabilities of the superradiant state seen in Fig.~\ref{fig:full_diagram} above $\omega>2E_0$ corresponds to the energies of two density-wave modes crossing at a line in parameter space. To deepen our understanding of why these crossings occur, and how they lead to instabilities, in this section we will first visualize the mode frequencies in Fig.~\ref{fig:modes}.  The modes can be separated into three families, between which crossings may occur. We will then more rigorously show that the instabilities can be  explained by crossings of two modes, by fitting the eigenvalues of the modes around an instability to the eigenvalues of a $2\times2$ matrix.

In Fig.~\ref{fig:modes}, we show the energy and decay rate spectrum of the lowest-energy excitations along cuts marked in Fig.~\ref{fig:full_diagram}. These are calculated from the same matrix $\mathcal{M}$ used for the linear stability analysis [Eq.~\eqref{eq:M_definition}]. 
That is, one takes the matrix $\mathcal{M}$ as defined above (see \resp{App.}{Appendix }~\ref{app:matrix_elements_calc} for details), and calculates its eigenvalues $\eta_\gamma = \Gamma_\gamma + i \omega_\gamma$.
The two panels in  Fig.~\ref{fig:modes} show the evolution of the real ($\Gamma_\gamma$) and imaginary ($\omega_\gamma$) parts of the eigenvalues as one varies the detuning $\omega$.
A similar mean-field treatment of collective modes is presented in Ref.~\cite{oztop_collective_2013} and summarized in Ref.~\cite{mivehvar_cavity_2021}.

The real part of an eigenvalue $\eta$ corresponds to the mode's energy (relative to the fixed point energy), and its imaginary part to its growth or decay rate (positive $\text{Im}(\eta)$ for growth, negative for decay, due to the $e^{-i\eta t}$ convention). 
We focus on low energy modes, which we expect to be primarily density wave excitations---there is a mode that is almost purely photonic at much higher frequencies $\sim\omega$. This is due to the large cavity width and detuning relative to the recoil frequency $\kappa\gg \omega_r$, $\omega\gg\omega_r$, implying a significant mismatch between density wave and photonic energies.

The figures also show the photonic participation factor~\cite{abed_participation_2000} of the mode $\gamma$
\begin{eqnarray}
    p^{(\gamma)}_\lambda=|l^{(\gamma)}_{\lambda}r^{(\gamma)}_\lambda|+|l^{(\gamma)}_{\lambda^*}r^{(\gamma)}_{\lambda^*}|,\label{eq:participation_factor}
\end{eqnarray}
with $l^{(\gamma)}$ and $r^{(\gamma)}$ being its corresponding left and right eigenvectors (i.e. $\mathcal{M} r^{(\gamma)} = \eta_\gamma r^{(\gamma)}$, and $l^{(\gamma)} \mathcal{M}  = \eta_\gamma l^{(\gamma)}$), normalized such that $\sum_i l^{(\gamma)}_i r^{(\gamma)}_i = 1$.
The subscript $\lambda,\lambda^\ast$ indicates the components of these eigenvectors corresponding to the cavity and conjugate cavity coordinate, $x\pm i y$ respectively, in terms of the vector in Eq.~\eqref{eq:Mevecdef}.
The low photonic weights confirm that the low energy modes are primarily of density-wave character. 

A notable feature of the low energy modes is that they separate into three different families: Modes excited in direction of the pump lattice, in direction of the cavity lattice, or in both directions. The derivation of this separation and classification of the modes is described in \resp{App.}{Appendix }~\ref{app:mode_class}. The families are distinguished by line-style in Fig.~\ref{fig:modes}.

Modes from different families may cross, since their energies depend on $\omega$ in different ways. In a Hermitian system, such a crossing of mode energies is generally avoided due to off-diagonal contributions. However, in non-Hermitian open systems, in general either the real or the imaginary part of the eigenvalue shows an avoided crossing, while the other part shows a crossing.  (The exception to this occurs at an exceptional point---a singularity in parameter space where both frequency and growth/decay rate cross---as discussed in e.g.~Refs.~\cite{eleuch_avoided_2013,heiss_repulsion_2000}.) 
We see this happening in Fig.~\ref{fig:modes}, where each intersection or avoided crossing of the energies also modifies the decay rates.  Where energy crossings occur, the decay rates show an anticrossing and vice versa. Close to these crossings, the decay rates and energies can also attract or repel each other.  If the repulsion of the decay rates is sufficiently strong, it leads one of the decay rates crossing zero, indicating an instability.
From the photonic participation factor, it can also be seen that the contribution of the dissipative cavity to the modes is increased at such resonances.

To further demonstrate that a two-mode crossing can sufficiently explain the resonances and instabilities, we fit the resonance at $\omega\approx \qty{91}{MHz}$ in Fig.~\ref{fig:modes}(a) to the spectrum of a $2\times2$ matrix,
\begin{eqnarray}
    \label{eq:fitmatrix}
    M(\omega)=\begin{pmatrix}
            \alpha_1(\omega+\beta_1) & \delta\\
            \delta & \alpha_2(\omega+\beta_2)
        \end{pmatrix},
\end{eqnarray}
with all parameters complex; see Fig.~\ref{fig:matrix_fit}. Note that $\delta$ is complex, and both terms involve $\delta$, not $\delta^\ast$, so the matrix is explicitly non-Hermitian~\footnote{We note that one could consider a more general matrix, with distinct elements $\delta_{1,2}$ on the off diagonal elements.  Such an expression would however have some degeneracies when considering only the eigenvalues of this matrix:  differences in the relevant modulus of $\delta_{1,2}$ can be compensated by changing the modulus of $\alpha_{1,2}$.  Similarly differences in phase between $\delta_{1,2}$ have no effect on the eigenvalues.  As such our ansatz is the most general non-degenerate ansatz for this $2\times 2$ problem.}. The fitting parameters found are given in the caption of Fig.~\ref{fig:matrix_fit}. We fit using weighted least-squares optimization, with the weights for the growth rates set to $10^3$ to compensate for their smaller magnitude. We find a good quality fit to this ansatz, confirming that a two-mode resonance in dissipative systems can lead to instabilities.

This type of resonance is a non-standard type of instability, not resulting from a Hopf bifurcation of a mode as described in Ref.~\cite{skulte_realizing_2024}, but instead from the amplification of an existing mode.

The resonances and their growth rates are hard to predict, since the type of crossing that takes place depends sensitively on the amplitude and phase of all matrix elements. Specifically, the phase of $\delta$ determines whether real or imaginary part cross. The phase of $\delta$ however depends on the relative phase of the eigenvectors, which do not show any simple trends. Notice, for example, the new resonance appearing at $P\approx0.72$ in Fig.~\ref{fig:full_diagram}, where the off-diagonal elements adjust to amplify the growth rate to above zero.

Such resonances and instabilities were not seen in previous theoretical works studying this system like Refs.~\cite{gao_self-organized_2023,nagy_self-organization_2008}, or experimentally in Ref.~\cite{baumann_dicke_2010} or ~\cite{klinder_dynamical_2015}. 
Theoretically, observing this phenomena requires including higher momentum states in two dimensions, while these cited theoretical works use one-dimensional treatments. 
It also likely only appears at high enough cavity decay rates, and, like bistability, is hard to observe in dynamics-focused treatments like Ref.~\cite{gao_self-organized_2023}.
When using experimental cavity loss rates, the instabilities have low growth rates for most choices of parameters and result in limit cycles at cavity field amplitudes quite close to their steady-state values (see Sec.~\ref{sec:UnstableDyn}). 

A similar effect resulting from mode resonances is described in Ref.~\cite{baur_bandstructure_2025}, which assumes two different polaritonic modes softening into two different superradiant phases, due to unequal forward and backward pumping resulting from an imperfect mirror. In the normal phase, these two polaritonic modes are predicted to synchronize in frequency at their energy crossing by coupling through the dissipative cavity, with one of the modes becoming unstable. The theoretical calculation there however relies on a simple approach to adiabatically eliminating  the cavity field, which cannot correctly predict whether or not an instability occurs. This would require an explicit inclusion of the cavity field into the dynamics as presented in our work. The synchronization is realized experimentally in Ref.~\cite{natale_synchronization_2025}, as well as potentially being the cause of the limit cycle excitations found in Ref.~\cite{dreon_self-oscillating_2022}. The experimental techniques used in these works might also be capable to detect the instabilities described here.

We conclude this discussion with a brief comment on the standard fate of limit cycles beyond mean field theory~\cite{Iemini2018BoundaryCrystals,fazio_many-body_2024}.
It is well known that the master equation always has at least one steady state solution.
For any large but finite size system, this corresponds to the phase-averaged behavior of the limit cycle.  That is, one considers different quantum trajectories showing limit cycles, but with different time offsets, and averages over them.
This leads to a stationary state.
Where the classical limit shows limit cycles, the full quantum solution will show a corresponding picture in the existence of a sequence Liouvillian eigenvalues with small decay rate and regularly spaced frequencies. As the system size goes to infinity, these decay rates tend to zero, leading to the possibility of infinitely long lived oscillations. 

\section{Dynamics in the unstable phases\label{sec:UnstableDyn}}

\begin{figure*}[htbp]
    \includegraphics[width=\textwidth]{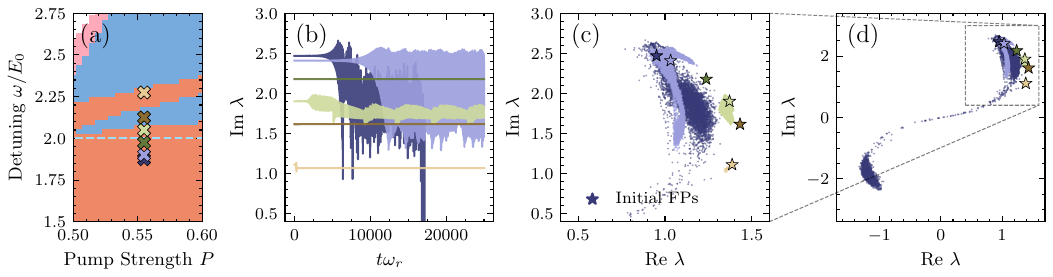}
    \caption{Dynamics of the cavity fields at a range of cavity frequencies crossing the boundary of the chaotic regime and some density wave resonances. \resp{}{(a) Color coded crosses on top of the phase diagram indicate the parameters of each simulation. (b) Imaginary part of the coherent light field $\lambda$ as a function of time. (c) Set of points visited by the dynamics focusing on the upper right quadrant of the complex plane. (d) Set of points visited by the dynamics in the full complex plane.}
    All simulations are initialized close to the superradiant fixed point, with a perturbation $\delta\lambda=10^{-5}$ In the chaotic phase (low $\omega$), the instabilities lead to chaotic behavior, with no periodicity. The fixed points then become pseudo-stable on the given timescale as $\omega$ is increased. A crossover between chaos and stability can be seen at $\omega/E_0\approx 1.9$ (second cross from bottom), where $\lambda$ does not switch sign and stays close to one attractor, but still behaves aperiodically. Meanwhile at density wave resonances, the dynamics converge towards different limit cycles, which can be complex with long periods as for $\omega/E_0\approx 2.05$, or simple oscillations as for $\omega/E_0\approx 2.25$. The fixed points between and after the density wave resonances are stable on all timescales.}\label{fig:field_ev}
\end{figure*}

This section explores the dynamics within the unstable phases identified in the phase diagram by using the time dependent solutions. We will first look at the large unstable region at $\omega\leq 2E_0$, and its transition to stability, then show that behavior in this region is consistent with a chaotic attractor.  Finally we will see that in the region we have so far labeled as chaotic, the dynamics can in fact converge to stable atomic superpositions.

The instability of steady state superradiance at $\omega\leq 2E_0$ was experimentally observed in Ref.~\cite{baumann_dicke_2010}, indicated by fluctuating photon counts. Our simulations reveal that the dynamics in this regime are not simple back-and-forth oscillations between a density wave and a normal phase, as might be naively assumed. Instead, they are generally aperiodic and characterized by apparently chaotic attractors. The theoretical existence of chaotic behavior and limit cycles in similar systems has been noted in 1D studies by Refs.~\cite{gao_self-organized_2023,tuquero_impact_2024,zwettler_cavity-mediated_2025}, with limit cycle-like dynamics also found experimentally in the latter.

In the following, we highlight a few key dynamical features of our 2D system. All dynamics are calculated using a Runge-Kutta integrator from the \emph{DifferentialEquations.jl} package~\cite{rackauckas_differentialequationsjlperformant_2017}. We adjust the solver's error tolerances to make sure that the wavefunction's norm is sufficiently conserved during the dynamics (i.e.~$\left| |\bm \phi(t_\text{final})|^2-|\bm \phi(t_\text{0})|^2 \right|=\left| |\bm \phi(t_\text{final})|^2-1\right|<10^{-4}$).

In Fig.~\ref{fig:field_ev}, we show the time evolution of the cavity field amplitude, and phase portraits of the cavity field, for different $\omega$. 
The values of $\omega$ used cross the instability threshold $\omega=2E_0$ and two density wave resonances. An apparently chaotic region and a crossover region confined to one attractor are visible for $\omega< 2 E_0$. At density wave resonances for $\omega> 2 E_0$, the dynamics converge to limit cycles of different periods. This behavior is fundamentally different from the chaotic regime, since even near the crossover to stability, the dynamics in the chaotic regime remain aperiodic.

\begin{figure*}[htbp]
    \centering
    \includegraphics[width=\linewidth]{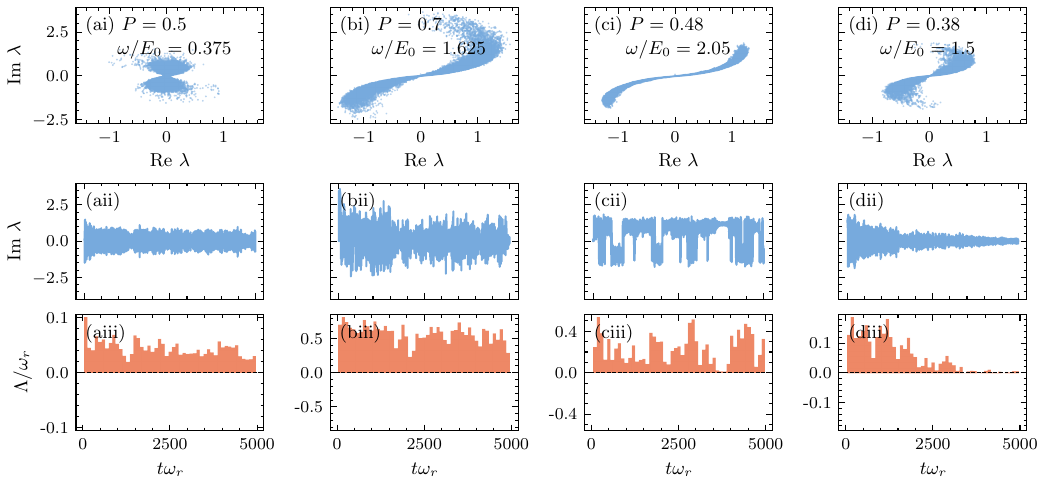}
    \caption{Dynamics and time dependent Lyapunov exponent (Eq.~\ref{eq:td_lyap}) of the light field at the parameters marked with diamonds in Fig.~\ref{fig:full_diagram}. Dynamics are initialized close to $\lambda=0$, except for (d) which is stable at $\lambda=0$, and is therefore initialized around its unstable fixed point at nonzero $\lambda$. \resp{Top row}{Row (i)}: phase space distribution of real and imaginary part of $\lambda$.  \resp{Middle row}{Row (ii)}: imaginary part of $\lambda$ over time. \resp{Bottom row}{Row (iii)}: time dependent Lyapunov exponents calculated over intervals $\Delta t=200/\omega_r$. The first three columns display different types of chaotic behavior, while the last column show convergence towards a stable attractor. Parameters are given in the panels.}. \label{fig:short_time_dynamics}
\end{figure*}

Inside the extended chaotic regime, there are qualitative differences between the dynamics at different parameters. We highlight a few types in Fig.~\ref{fig:short_time_dynamics}. To show the trajectories are chaotic, we also show the time-dependent largest Lyapunov exponent, which we elaborate on here.

Chaotic behavior in dynamical systems is intuitively defined by sensitivity to initial conditions: Two trajectories with slightly different initial conditions, no matter how similar, will eventually diverge away from each other. This is in contrast to convergent behavior, where both trajectories converge to the same fixed point and therefore also converge towards each other, or limit cycles, where the distance stays constant. 

\begin{figure}[htbp]
    \centering
    \includegraphics[width=\linewidth]{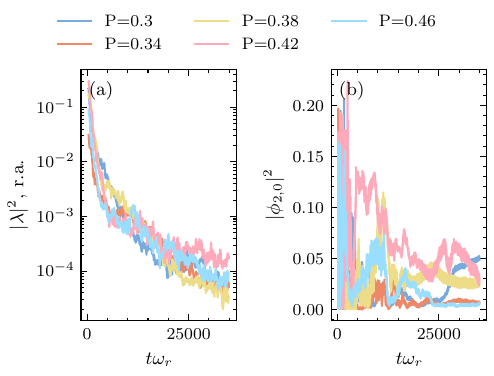}
    \caption{\resp{}{(a) }Evolution of the light field and \resp{the}{(b)} population of the $\phi_{2,0}$ momentum state\resp{(inset) }{} in the SAS phase for different $P$ at $\omega/E_0=1.25$. While the light field slowly converges towards zero, nonzero momentum atomic states remain occupied, and the system remains in an oscillatory state. This behavior is seen in the normal, normal+, and dynamical phase (see the phase diagram in Fig.~\ref{fig:full_diagram} for the extent of where such behavior is stable).}\label{fig:SAS_convergence}
\end{figure}

This intuition can be formalized using Lyapunov exponents~\cite{datseris_nonlinear_2022}. Given a dynamical system $\dot{\bm x}=f(\bm x)$, we solve for two initial conditions $\bm x_1(t_0)=\bm x_0$ and $\bm x_2(t_0)=\bm x_0+\bm \delta_0$, with $\bm \delta_0$ representing a small perturbation of the initial condition. We then track the distance between the trajectories, which we call $\bm \delta(t)=\bm x_2(t)-\bm x_1(t)$. For a chaotic system, $|\bm \delta(t)|$ can be oscillatory over short timescales, but needs to grow on average for the trajectories to diverge from each other. Eventually, the trajectories should become completely independent due to the nonlinearity of the equations, and $|\bm \delta(t)|$ saturates. Since the initial divergence is what classifies the chaos, we take an infinitesimal $|\bm \delta_0|\rightarrow 0$. This will not be affected by nonlinearities between the two trajectories, and the quotient $|\bm \delta(t)|/|\bm \delta_0|$ will still show growth or decay. Additionally, in a truly chaotic system, the chaotic behavior is ongoing, and the system never converges to a fixed point or limit cycle, and we can therefore take the limit $t\rightarrow\infty$. This limit also ensures that the oscillations of $\bm \delta(t)$ on short timescales do not play a role. The largest Lyapunov exponent is therefore defined as ~\cite{datseris_nonlinear_2022}:
\begin{eqnarray}
    \Lambda=
    \lim_{t\rightarrow\infty}
    \lim_{|\delta_0|\rightarrow0}
    \frac{1}{t}\ln \frac{|\bm \delta(t)|}{|\bm \delta_0|}.
    \label{eq:lyapunov}
\end{eqnarray}
It measures whether an infinitesimally perturbed trajectory will converge towards, or diverge away from, the original trajectory, with positive $\Lambda$ indicating chaotic behavior. The exponents shown in Fig.~\ref{fig:short_time_dynamics} are calculated using the \emph{ChaosTools.jl} package~\cite{datseris_nonlinear_2022}. 

When calculated numerically, the limit $t\rightarrow\infty$ cannot be taken, and the trajectories classified as chaotic can include transient chaos, which eventually converges to a fixed point or limit cycle. Indeed, Lyapunov exponents taken over short times will be positive for trajectories that converge to a stable configuration eventually. Since transient chaos over experimentally relevant timescales is also of interest in this system, in Fig.~\ref{fig:short_time_dynamics} we show time-dependent Lyapunov exponents of the form
\begin{eqnarray}
    \Lambda(t)=\lim_{|\delta_0|\rightarrow0}\frac{1}{\Delta t}  \ln \frac{|\bm \delta(t+\Delta t)|}{|\bm \delta(t)|}\label{eq:td_lyap}
\end{eqnarray}
This describes the divergence of the two trajectories initialized at infinitesimal distance $\bm\delta_0$ between times $t$ and $t+\Delta t$. $\Delta t$ here defines a timescale over which the trajectories are evolved, before evaluating the quotient $|\bm \delta(t+\Delta t)|/|\bm \delta(t)|$. The limit $\Delta t\rightarrow\infty$ recovers the time-independent Lyapunov exponent, and the limit $\Delta t \rightarrow 0$ corresponds to a linearized stability analysis around the trajectory at each point in time. We choose $\Delta t$ large enough such that short time oscillations of the exponents are averaged out, but ensuring that the decay of transient chaos and the impact of attractor switching can still be seen.

Practically, to keep $\delta$ infinitesimal, we use the periodic resetting technique described in Ref.~\cite{datseris_nonlinear_2022} and sum $\Lambda$ over the intermediate resetting times. For the dynamics over experimentally relevant timescales shown in Fig.~\ref{fig:short_time_dynamics}, the Lyapunov exponents confirm that behavior observed can be identified as chaotic. While in the quasinormal phase (Fig.~\ref{fig:short_time_dynamics}(d)), the dynamics converge to the stable attractor with Lyapunov exponents decaying to zero, the dynamical phases (Fig.~\ref{fig:short_time_dynamics}(a-c)) display different chaotic behaviors, interpolating between the $\lambda=0$ attractor at $\omega\ll E_0$ and switching between unstable attractors at $\omega\approx 2E_0$.

Inside the region of stable atomic superpositions---particularly where this overlaps with the unstable region at $\omega< 2E_0$ and at smaller pump strength---the light field converges to $\lambda=0$ on longer timescales, while the atoms remain in an oscillatory superposition. Since the excitations around the superpositions often have small decay rates, as elaborated on in Sec.~\ref{sec:SAS}, the light field only decays slowly once the dynamics revolve around the SAS state (see Fig.~\ref{fig:SAS_convergence}).

In Ref.~\cite{gao_self-organized_2023} there is a small limit cycle phase between the superradiant and chaotic regime at $\omega\ll E_0$. We were not able to find such a phase using our parameters, since our dynamics turn chaotic right after the superradiant fixed point becomes unstable. We assume this phase may appear at lower cavity decay $\kappa$, as it is observed experimentally for a recoil-resolved ($\kappa<\omega_r$) cavity in Ref.~\cite{skulte_realizing_2024}.

\subsection{\resp{}{Effects of Noise on Limit Cycles}}

\resp{}{
Beyond the classical limit, the full quantum density matrix of the system will generically show a steady state, even in these limit cycle phases.  
This is because---as discussed earlier---the full quantum density matrix involves an ensemble average over limit cycles with different phases (i.e. time-origins of the limit cycle).
This can be thought as phase diffusion arising from quantum noise.
Such behavior has in fact been studied a number of times for the Dicke model~\cite{Emary2003ChaosModel,mondal_transient_2025}.
Furthermore, the effects of noise has been analyzed for a one-dimensional model~\cite{gao_self-organized_2023} (i.e.\ neglecting atomic dynamics in the pump direction).  That work showed that in such a treatment the chaotic, limit cycle, and SAS behavior were all stable to cavity loss noise. 
This work also studied the effects---within mean-field theory---of weak contact interactions between atoms, and again found no qualitative change to the behavior in any phase.
}

\resp{}{Noise however may have an impact on the experimental detectability of the polariton resonances, as it could wash out limit cycle oscillations. 
The standard way of incorporating photon field shot noise within a mean-field treatment is through a stochastic Langevin term~\cite{gardiner_quantum_2000}, which has already been applied to a similar system in Ref.~\cite{tuquero_impact_2024}.  In this section we evaluate the impact of such noise on the polaritonic resonances  by simulating the stochastic differential equation
\begin{eqnarray}
    \frac{d \phi_{n, m}}{d t}&=&  -i \omega_r\left[\left(n^2+m^2\right) \phi_{n, m}-|\lambda|^2 \sum_{s \in \pm 1} \phi_{n, m+2 s}\right.\label{eq:EoM_atoms_noise} \\
    &-&\left.P\left(\lambda+\lambda^*\right) \sum_{s, s^{\prime}\in \pm 1} \phi_{n+s, m+s^{\prime}}-P^2 \sum_{s \in \pm 1} \phi_{n+2 s, m} \right],\nonumber \\
    \frac{d \lambda}{d t}&=& -(\kappa+i \omega) \lambda+i E_0 \lambda \sum_{\substack{n,m\\s  \in \pm 1}} \phi_{n, m+2 s}^* \phi_{n, m} \nonumber\\
    & +&i E_0 P \sum_{\substack{n,m\\s, s^{\prime}  \in \pm 1}} \phi_{n+s, m+s^{\prime}}^* \phi_{n, m} + i\sqrt{\frac{\kappa E_0}{N_a \omega_r}}\xi(t).\label{eq:EoM_cavity_noise}
\end{eqnarray}
where $N_a$ is the atom number and $\xi$ is a Wiener noise term defined by $\langle \xi(t)\xi(t')\rangle=\delta(t-t')$. We set $N_a=10^5$ in accordance with the experiment described in Ref.~\cite{baumann_dicke_2010}. The prefactor of the noise term results from the conversion of photon number to the dimensionless light field $\lambda$ defined in Eq.~\eqref{eq:dimless_definitions}. 
As one might expect, the noise term becomes zero in the $N_a\rightarrow \infty$ thermodynamic limit. }

\begin{figure}[htpb]
    \centering
    \includegraphics[width=\linewidth]{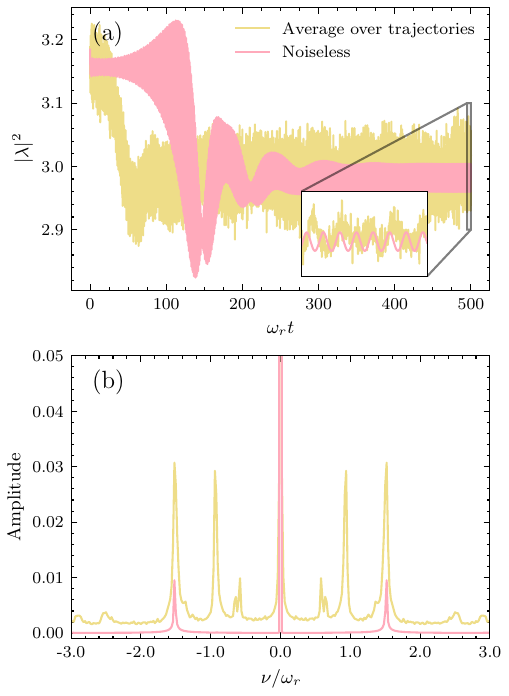}
    \caption{\resp{}{Average over 64 noisy trajectories from simulating Eqs.~(\ref{eq:EoM_atoms_noise}) and (\ref{eq:EoM_cavity_noise}), compared to the noiseless Eqs.~(\ref{eq:EoM_atoms}) and (\ref{eq:EoM_cavity}). In, (a) the average is taken over all trajectories at each time $t$, while for (b) a Fourier transform is taken over the last $\Delta t=50\omega_r$ of each trajectory, and the average over all trajectories is taken at each frequency. In the frequency domain, the characteristic frequency of the limit cycle is clearly visible, and its amplitude increased compared to the noiseless trajectory.}}
    \label{fig:noisy_dynamics}
\end{figure}

\resp{}{
We simulate $N_{\text{traj}}=64$ different trajectories starting from the unstable superradiant fixed point. We monitor the time evolution of $|\lambda|^2$ vs time, corresponding to the intracavity photon count, in order to remove the fast phase evolution of the cavity field. We show the ensemble average of $|\lambda|^2$ both in time and frequency domain in Fig.~\ref{fig:noisy_dynamics}. For the time average in Fig.~\ref{fig:noisy_dynamics}(a), this means $(|\lambda|^2)_\text{avg}(t)=\sum_n |\lambda_n(t)|^2/N_{\text{traj}}$, with the sum running over the individual trajectories. For the spectrum in Fig.~\ref{fig:noisy_dynamics}(b) $|\lambda^2|_\text{avg}(\nu)=\sum_n \int_{t_{eq}}^{t_f} e^{i\nu t} |\lambda_n(t)|^2 dt/N_{\text{traj}}$, where $t_{eq}$ is a time chosen so that the mean field model has stopped transient evolution, and $t_f$ the final time. While the oscillations of the trajectories become dephased from each other, leading to little oscillation in the mean at each time, the photon number in each realization still oscillates with the characteristic frequency of the limit cycle. However, there are additional frequencies corresponding to other soft polariton modes, the main one being the lowest energy mode in Fig.~\ref{fig:modes}(a) which has some photonic participation. 
This suggests that experimentally, the resonance would not appear as a pure limit cycle, but may take the form of a more complex dynamical state.
However, there should still be clear signatures of the instability both through the additional frequencies appearing in the dynamics, as well a the change to the average photon number.
Interestingly, for noisy trajectories, the convergence towards the ``limit cycle'' state in time is accelerated compared to the pure mean field simulations, and amplitude corresponding to the cycle frequency increased. This is consistent with the analysis of Ref.~\cite{tuquero_impact_2024}, which notes that  noise generally increases the domain of limit cycle phases.
}

\section{Stable Atomic Superpositions\label{sec:SAS}}
\begin{figure}[htbp]
    \centering
    \includegraphics[width=\linewidth]{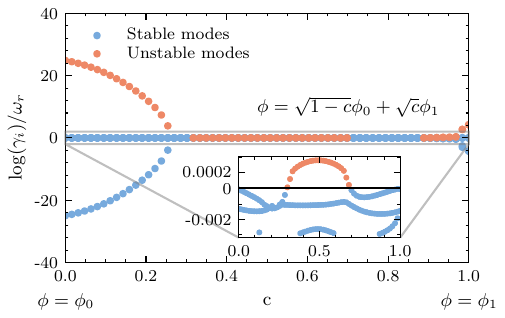}
    \caption{The eigenvalue spectrum of the monodromy matrix $\{\gamma_i\}=\text{eig}(U(T))$ (Eq.~\eqref{eq:floquet}) for a limit cycle involving a superposition of the ground state $\bm\phi_0$ and the first cavity-direction excited state $\bm\phi_1$. 
    The logarithm is plotted to convert the values to growth rates, with positive growth rates indicating instability, shown in orange, and stable modes shown in blue. Trivial, atomic only modes with eigenvalue $|\gamma_i|=1$ are not shown. Towards $c=0$ or $c=1$ (pure excited or ground state), bifurcations leading to instabilities appear, while for superpositions, there are stable regions. The inset shows the region close to $F=1$ on a symlog scale. Parameters are $P=0.55$, $\omega/E_0=0.6875$. 
    }\label{fig:floquet_superpositions}
\end{figure}

In this section, we discuss the stability of  atomic superposition states.  As discussed in Sec.~\ref{sec:UnstableDyn}, we found that some dynamics converge to atomic superpositions which oscillate but exhibit vanishing cavity field. This convergence has also been observed in 1D in Ref.~\cite{gao_self-organized_2023}, there identified as a separate phase.  We seek to further understand that behavior here.

The atomic superpositions observed in simulations only occupy atomic momentum states with even $n$ and $m$. For such states, the cavity coupling term $\sum_{s,s'\in\pm1} \phi_{n+s,m+s'}^*\phi_{n,m}$ in Eq.~\eqref{eq:EoM_cavity} vanishes. This corresponds to the atoms being ordered with a periodicity of $1/2q$, resulting in purely destructive interference of scattered pump photons, such that the cavity field can satisfy $\lambda=0$ at all times.

As the atomic momentum states occupied in this superposition have different energies, the state is oscillatory.  In this sense, these states are similar to limit cycles.  However,  as we will see below, these atomic superposition states can form a continuum of different stable states, and so differ from the standard definition of an isolated limit cycle.

Based on the above, we may note that any state with $\lambda=0$ and a superposition of even atomic momentum states is a solution of equations of motion.  However, as with the normal state (which can be considered a limiting special case of these atomic superpositions), such states are not always stable.  We thus next discuss their stability.

The linear stability analysis presented in Sec.~\ref{sec:LinearStability} only considers the stability of fixed points.  
To investigate the stability of atomic superpositions, we treat these superpositions in the same way as is used to determine the stability of  periodic states such as limit cycles, by employing Floquet theory.
(Not all atomic superpositions are necessarily periodic, as discussed below, however we can apply the following to consider the subset of atomic superpositions which are periodic.)
An overview of the Floquet analysis method is provided below, with further details available in Refs.~\cite{grimshaw_nonlinear_2017,klausmeier_floquet_2008}.
For a known periodic trajectory $x_0(t)$, we write down a perturbed trajectory as 
\begin{equation}
    x(t)=x_0(t)+\delta x(t).
\end{equation}
$x_0(t)$ follows the periodic dynamics, and $\delta x(t)$ describes the evolution of a perturbation around it. When inserting this into the equations of motion, we can linearize in $\delta x(t)$ and get the time-dependent linear equation:
\begin{equation}
    \frac{d \delta x(t)}{dt}=i\mathcal{M}[x_0(t)]\delta x(t).\label{eq:pert_evolution}
\end{equation}
Since $\mathcal{M}(t)=\mathcal{M}[x_0(t)]$ is periodic by construction, we can treat Eq.~\eqref{eq:pert_evolution} using Floquet analysis. Consider an initial perturbation $\delta x(0)$. Eq.~\ref{eq:pert_evolution} has a general solution $x(t)=U(t)x(0)$, with the matrix solution $U(t)$ acting as a time evolution operator of the linearized system. The matrix $U(t)$ can be explicitly constructed by taking small time steps $\Delta t$:
\begin{equation}
    U(t)=e^{-i\mathcal{M}(t-\Delta t)\Delta t} e^{-i\mathcal{M}(t-2\Delta t)\Delta t}\dots e^{-i\mathcal{M}(0)\Delta t}.\label{eq:monodomy_numerical}
\end{equation}
Numerically, this is the formula used, with some small but finite $\Delta t$. In the continuum limit, it corresponds to the time ordered integral
\begin{equation}
    U(t)=\mathcal{T} e^{-\int_0^t i\mathcal{M}\left(t'\right)t'dt'}.\label{eq:floquet}
\end{equation}
For the stability analysis of a periodic solution, we thus consider the monodromy matrix $U(T)$, where $T$ is the limit cycle period. If $U(T)$ has any eigenvalues $\gamma_i$ for which $|\gamma_i|>1$, there is an initial perturbation vector $x_i$ that will grow each period, since $||U(T)x_i||=||\gamma_i x_i||>||x_i||$, and the periodic solution is considered unstable. Otherwise, every perturbation will decrease in magnitude, and the solution is stable. If a mode has $\gamma=1$, the stability is marginal. 
Stability is thus determined by whether $\gamma_{\text{max}}\equiv\max_i(|\gamma_i|)$ is greater than one or not.

Numerically, calculating the integral in a stable way involves repeated matrix exponentials across time steps. This is computationally costly but still feasible, since at $\lambda=0$, only the momentum states coupling to $x_0(t)$ need to be included.
To clearly determine the stability, we filter out some purely atomic modes with Floquet multiplier $\gamma=1$ by discarding states that have no photonic participation. While marginal oscillations could therefore remain in the atomic wavefunction, this guarantees that the cavity field converges to zero close to the periodic solutions, which is the phenomenon we are investigating.

We first consider periodic atomic superposition described by
\begin{eqnarray}
    &\bm\phi_{AS}(t)&=\sqrt{1-c}\bm\phi_0+e^{-4i\omega_r t}\sqrt{c}\bm\phi_1,\\
    &\lambda(t)&=0,
\end{eqnarray}
where $\bm\phi_0$ corresponds to the atomic ground state, and $\bm\phi_1$ to the first excited state in the cavity direction.  
Note that in writing this expression we have gauged out the time evolution due to the ground state energy, i.e. a common factor of $e^{-i \epsilon_0 t}$ multiplying $\phi_{AS}(t)$.
Note also that both states $\bm{\phi}_{0}$, $\bm{\phi}_1$  are calculated in the presence of the pump lattice.  While the pump lattice shifts the energies of the atomic density waves, the energy difference between $\bm\phi_0$ and $\bm\phi_1$ remains exactly $4\omega_r$ as we are working at $\lambda=0$. 

We find that, while the excited state by itself---i.e.~the limit $c=1$---is always unstable, there are regions of the superposition coefficient $c<1$ for which these superpositions can be stable, creating a continuum of possible stable periodic solutions (see Fig.~\ref{fig:floquet_superpositions}). 
Notably, there are parameters where the normal state $c=0$ is unstable, but periodic solutions with $0<c<1$ are stable.
In the phase diagram (Fig.~\ref{fig:full_diagram}), we shade the region in which these stable atomic superpositions (SAS) can occur. These solutions are bistable with the stationary states if any exist. If one considers these SAS states as generalizations of the normal state, they extend the region over which the normal state is stable.

Our analysis for the whole phase diagram only includes two atomic momentum states, since including more would make a scan prohibitively computationally expensive. Including higher order excited cavity states can extend the stable region. We check this for a fixed $\omega$ by minimizing the largest Floquet multiplier of the periodic solutions
\begin{eqnarray}
    &\bm\phi_{AS}(t)&=\sqrt{1-\sum_{i=1}^{N} c_i}\bm\phi_0\\
    &&+e^{-4i\omega_r t}\sqrt{c_1}\bm\phi_1+e^{-16i\omega_r t}\sqrt{c_2}\bm\phi_2+\ldots,\nonumber\\
    &\lambda(t)&=0, \label{eq:multi_state_LC}
\end{eqnarray}
Here $\bm \phi_0$ again denotes the ground state (calculated in the presence of the pump lattice), while the excited cavity states $\bm \phi_i$ are then shifts of the original ground state in the cavity direction by two times the recoil momentum, i.e. $(\bm \phi_i)_{n,m}=(\delta_{2i,m}+\delta_{-2i,m})(\bm \phi_0)_{0,n}/\sqrt{2}$. 
We restrict to atomic excitations in the cavity direction because the energy splitting between these states is then an integer multiple of $\omega_r$, leading to a periodic solution.
Excitations in the pump direction have incommensurate energies, lead to an aperiodic solution which may be stable, but is not amenable to Floquet analysis.
We do, however, observe numerical convergence to states including excited pump states when simulating dynamics, suggesting such a $\lambda=0$ aperiodic solution may be possible.

We can determine stability by numerically minimizing $\gamma_{\text{max}}=\max_i \gamma_i$ over the coefficients $c_j$, using a Nelder--Mead optimizer\cite{nelder_simplex_1965,gao_implementing_2012}. We find that the stable region is significantly extended by including a third and fourth state, but with no significant change found beyond that.
However, with a larger state space to search, the computational cost of determining the least unstable configuration becomes significantly higher, and pockets of stability may be missed for these more complex states.
It remains an open question what the real extent of potential stable superpositions is, or if they even extend to infinity given enough states in the superposition. 
Due to relatively long times per function call, a global minimum search is not currently feasible.

This feature bears some resemblance to matter wave superradiance ~\cite{inouye_superradiant_1999,kesler_steering_2014} where atoms are irreversibly scattered into higher momentum states through superradiant light interaction. This phenomenon is however usually associated with sharp spiking of the photon count instead of long decay times, and does not generally require a cavity.
Reference~\cite{klinder_observation_2015} reported cavity assisted matter wave superradiance experimentally at negative $\omega$, resulting in a final stable state of occupied even-order momentum states at zero cavity field as found here. With Ref.~\cite{baumann_dicke_2010} also finding the photon count spiking and decaying to zero at positive $\omega$, these dynamical phenomena may have a similar physical origin of stable superpositions existing in these regimes, with no stable superradiant state.

\section{\label{sec:conclusion}Conclusions and Outlook}

We have shown that the transversely pumped BEC in a cavity displays a remarkably rich phase diagram, with several features that are absent when considering either the Dicke-model approximation or one-dimensional simulations.  
These features all exist in regimes that are possible to access experimentally.

By analyzing the steady states of the system, we find regions of bistability between normal and superradiant phases and a tricritical point, which result from the inclusion of higher cavity momentum states. This bistability should be replaced by a first-order transition in a full quantum treatment, but experimental observation of bistability should still be possible due to metastability of the two states.

We also observe instabilities of superradiance due to density-wave resonances, resulting in limit cycles. While with the given setup and parameters, these instabilities remain weak and hard to detect experimentally, they seem to be a general feature of 2D cavity systems and might be more relevant in other setups.

Stable atomic superpositions have been identified as a continuum of stable periodic solutions at vanishing cavity field. This is a curious feature of this mean-field analysis, and its physical significance beyond mean-field is unknown to us. Experimentally, these superpositions might be accessible, so this extension of the normal phase for superpositions could be probed.

We have commented in various places about what would be expected in a full quantum treatment of the model we study, based on established results for the fate of bistability, limit cycles, and chaotic attractors in full quantum treatments.  Simulating the full quantum dynamics of the current model would be computationally challenging, and likely limited to small numbers of atoms.
A further challenge would be to consider how a full (beyond mean-field) treatment might be affected by processes beyond the model we consider, such as atom-atom interactions, or other sources of dephasing. In particular, it would be interesting to study the fate of the stable atomic superposition phase in the processes.

Their variety of behaviors makes coupled atom-cavity systems an outstanding model to study the physics of driven-dissipative quantum systems. \resp{}{Our analysis of the density wave ordering in the single mode case is also of  relevance for more complex cavity setup, like multimode~\cite{gopalakrishnan_emergent_2009,gopalakrishnan_frustration_2011,kollar_adjustable-length_2015,kollar_supermode-density-wave-polariton_2017,vaidya_tunable-range_2018} or multiple cavity~\cite{leonard_supersolid_2017,gopalakrishnan_intertwined_2017} setups. Through our methods, we can \resp{clearly}{} find which regions of the phase diagram really have a second order transition, and whether the system can be approximated as a Dicke model or one instead must consider the impact of higher momentum states}. Further studies could be both experimental, probing the detectability of the new features, or theoretical, extending the model to include more quantum effects, other parameter regimes, or fermionic atoms.

\section{Data availability}
\resp{}{The data that support the findings of this article are openly available\cite{data_availability}.}

\begin{acknowledgments}
    We are grateful for helpful discussions with A.~Sheikhan.
    Results presented in this manuscript were calculated using the University of St Andrews High Performance Computing Facility, Hypatia.
    J.~M.~acknowledges funding from a  ``St Andrews and Bonn Global PhD'' scholarship from the University of St. Andrews and Rheinische Friedrich-Wilhelms-Universit\"{a}t Bonn.
    M.~L.~S.~acknowledges support from EPSRC (Grant No. EP/M506631/1).
    S.B.J and C.K acknowledge support from
    the Deutsche Forschungsgemeinschaft (DFG, German Research Foundation) under Project No. 277625399-TRR 185 OSCAR (“Open System Control of Atomic and Photonic Matter”, B4), No. 277146847-CRC 1238 (“Control and dynamics of quantum materials”, C05), CRC 1639 NuMeriQS (“Numerical Methods for Dynamics and Structure Formation in Quantum Systems”) – project No. 511713970, and under Germany’s Excellence Strategy – Cluster of Excellence Matter and Light for Quantum Computing (ML4Q) EXC 2004/1 – 390534769. 
    J.K.~acknowledges support from EPSRC (Grant No. EP/Z533713/1)
\end{acknowledgments}

\appendix
\section{Classification of the mode types\label{app:mode_class}}
The low energy modes shown in Fig.~\ref{fig:modes} seem to separate into those that are excited in either the cavity direction, the pump direction, or both. To understand the origin of this splitting and correctly classify the modes, we here introduce a number of approximations. \\
The total atomic potential in real space is given by
\begin{align}
    V(x,z)&=V_\text{cavity}(z)+V_\text{pump}(x)+V_\text{cross}(x,z)\label{eq:full_pot}\\
    V_\text{cavity}(z)&=\omega_r|\lambda|^2\cos(2z)\label{eq:full_cavity_pot}\\
    V_\text{pump}(x)&=\omega_rP^2 \cos(2x)\label{eq:full_pump_pot}\\
    V_\text{cross}(x,z)&=\omega_rP(\lambda+\lambda^*)\cos(x)\cos(z).\label{eq:full_cross_pot}
\end{align}
Where the real space coordinates $x$,$z$ are given in units of $1/q$ such that $\omega_r$ is the only dimensionful number. This potential is not separable due to $V_\text{cross}$. Since the modes anyway seem to approximately separate in $x$ and $z$ direction, we consider the ansatz
\begin{eqnarray}
    \phi(x,z)=\phi_x(x)\phi_z(z).
\end{eqnarray}
We also assume $|\lambda|\gg P$, which is given in the relevant region of parameter space. Writing down the Schroedinger equation for the given potential and ansatz, we find
\begin{eqnarray}
    \frac {E}{\omega_r} \phi_x(x)\phi_z(z)&=&-\phi_x \partial^2_z \phi_z  -\phi_z\partial^2_x\phi_x\nonumber \\
    &+&\left[|\lambda|^2\cos(2z)\right. \nonumber\\
    &+&P(\lambda+\lambda^*)\cos(x)\cos(z)\nonumber\\
    &+&\left. P^2 \cos(2x)\right]\phi_x(x)\phi_z(z)
\end{eqnarray}

For the wavefunction $\phi_z$, the dominant potential is $\propto |\lambda|^2$, so we can neglect $V_\text{cross}$ and $V_\text{pump}$, multiply by $\phi^*_x$, integrate both sides over $x$ and write
\begin{eqnarray}
    E_z\phi_z=\omega_r[-\partial^2_z +|\lambda|^2 \cos(2z)]\phi_z\label{eq:SE_cavity}
\end{eqnarray}
which is the one-dimensional Schroedinger equation we solve when finding the mode energies in cavity direction. The resulting eigenmodes are the Mathieu functions.

In $x$, or pump direction, the dominant potential is instead $\propto P(\lambda+\lambda^*)$. Multiplying by $\phi_z^*$ and integrating both sides over $z$, we get
\begin{equation}
    E_x\phi_x=\omega_r[-\partial_x^2 +P(\lambda+\lambda^*)\cos(x)\int dz \cos(z)|\phi_z|^2]\phi_x
\end{equation}

\begin{figure}
    \centering
    \includegraphics[width=\linewidth]{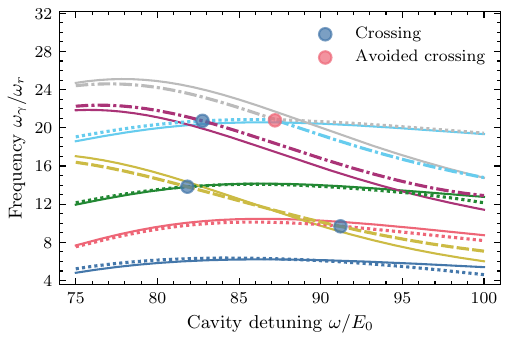}
    \caption{The mode energies from Fig.~\ref{fig:modes}(a), dashed, dotted and dashdotted, compared to the approximated energies computed from Eqs.~\ref{eq:SE_cavity} and~\ref{eq:SE_pump}, solid lines.}
    \label{fig:separabel_approx}
\end{figure}

The integral over $z$ can not be completely evaluated due to the remaining $\cos z$ in the dominant potential. We therefore have to consider the solutions of Eq.~\ref{eq:SE_cavity}. This is analogous to a Born-Oppenheimer approximation, where the solution to a fast potential determines an effective slow potential. If the confinement proportional to $|\lambda|^2$ is sufficiently strong, the low energy solutions $\phi_z$ will be localized close to the antinodes of $\cos(z)$. Therefore, we can approximate
\begin{eqnarray}
    \int dz |\phi_z|^2 \cos(qz)  \approx \int dz |\phi_z|^2\times 1=1.
\end{eqnarray}
Note that this approximation is only valid as long as $\phi_z$ is in a sufficiently low-energy state. To determine the possible modes in $x$, or pump direction, we therefore solve the Schroedinger equation
\begin{eqnarray}
     E_x\phi_x=\omega_r[-\partial_x^2 +P(\lambda+\lambda^*)\cos(x)]\phi_x\label{eq:SE_pump}
\end{eqnarray}
which also results in states shaped like the Mathieu functions.\\
The family of each mode $\mathcal{V}$ from the full linear stability analysis is then determined by first calculating the lowest eigenvalues $E_{x,i}$, $E_{z,j}$ of both Eqs.~\ref{eq:SE_cavity} and \ref{eq:SE_pump}. We also construct the mixed mode energies as $E_{m,ij}=E_{x,i}+E_{z,j}$. By comparing these approximate energies to those of the exact solutions, we can assign a family to each exact solution depending on its closest approximate solution. Note that this can fail, especially when classifying the modes close to a mode crossing, and needs to be visually confirmed. In Fig.~\ref{fig:separabel_approx}, we compare the approximated mode energies (solid lines) to the energies calculated from the full model (dashed and dotted lines). 
\section{Effects of truncation of the atomic state}\label{app:error_analysis}
\begin{figure*}
    \centering
    \includegraphics[width=\linewidth]{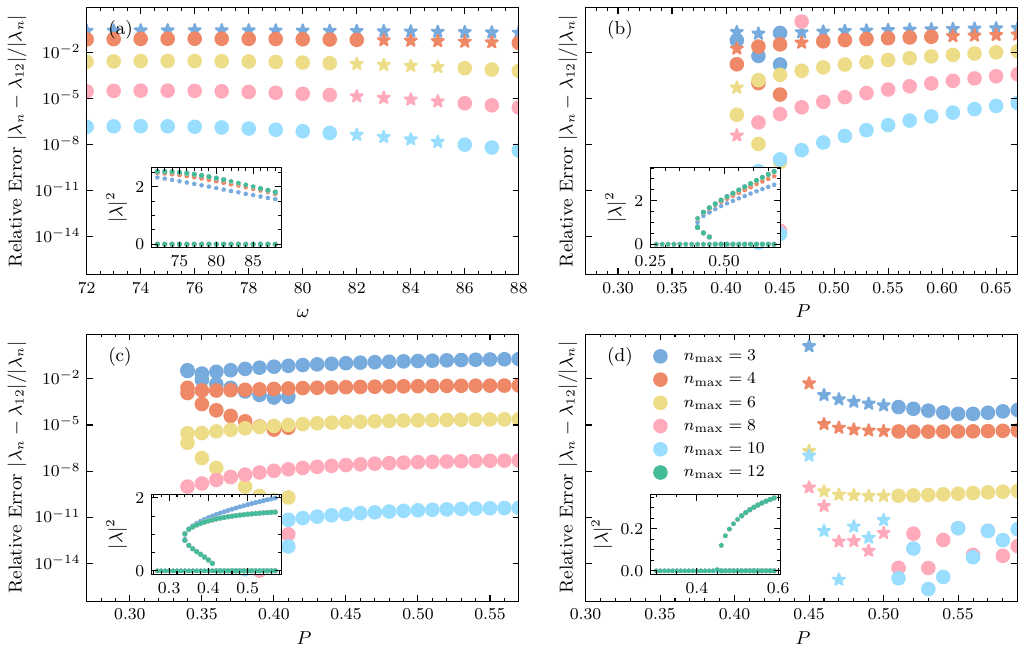}
    \caption{Analysis of the error of the steady state photon field $\lambda$ along multiple slices of the phase diagram, depending on the number of momentum states $n_{\max}=m_{\max}$ included in the calculation. Parameters are (a) $P=0.53$, (b) $\omega=79$, (c) $\omega=56$, (d) $\omega=9$. Stable fixed points are indicated by a star, unstable ones by a circle. We compare the value of $\lambda_n$ for each $n_{\max}$ with $n_{\max}=12$, and plot the relative error $\Delta\lambda=\frac{|\lambda_n-\lambda_{12}|}{|\lambda_n|}$. We only compute this for the nonzero fixed points, since $\lambda=0$ is exact for any $n$. The inset shows the fixed points themselves for each $n$.  We find stability converging relatively fast. The maximum relative error for $n_{\max}=10$ is $10^{-5}$ in (b). We can therefore assume the relative error of $n_{\max}=12$ compared to the fully converged value to be less than $10^{-5}$.}
    \label{fig:error_analysis}
\end{figure*}

To show that the number of atomic states $n_{\max}=m_{\max}=12$ included is sufficient to make quantitative predictions, we analyze the convergence of the superradiant steady state value of $\lambda_0$ along different cuts of the phase diagram, see Fig.~\ref{fig:error_analysis}. We find a maximum possible error $|\Delta\lambda| < 10^{-5}$, which corresponds to much less than a single photon, and is therefore sufficiently accurate.
\section{Elements of the linear stability matrix \texorpdfstring{$\mathcal{M}$}{}}\label{app:matrix_elements_calc}
We here present the elements of the linear stability matrix $\mathcal{M}$ defined in Eq.~\ref{eq:M_definition}. We name the indices of the matrix by the quantity their coordinate corresponds to. The indices $a_{n,m}$ and $b_{n,m}$ correspond to $\delta\phi_{n,m}$ and its complex conjugate respectively, as do $x$ and $y$ correspond to $\delta\lambda$ and its complex conjugate.
\begin{equation}
        \begin{aligned}
            \mathcal{M}_{a_{n,m},a_{n',m'}}&=\omega_r\left[\right.(n^2+m^2-\epsilon_0) \delta_{n,n'}\delta_{m,m'}\\
            &-|\lambda_0|^2 \sum_{s\in\pm 1} \delta_{n,n'}\delta_{m,m'+2s} \\
            & -P\left(\lambda_0+\lambda_0^*\right) \sum_{s, s^{\prime} \in\pm 1} \delta_{n,n'+s}\delta_{m,m'+s}\\ 
            &- P^2 \sum_{s\in\pm 1} \delta_{n,n'+2s}\delta_{m,m'}\left.\right]
        \end{aligned}
    \end{equation}
    \begin{equation}
        \begin{aligned}
            \mathcal{M}_{b_{n,m},b_{n',m'}}&=-\omega_r\left[\right.(n^2+m^2-\epsilon_0) \delta_{n,n'}\delta_{m,m'}\\
            &-|\lambda_0|^2 \sum_{s\in\pm 1} \delta_{n,n'}\delta_{m,m'+2s} \\
            & -P\left(\lambda_0+\lambda_0^*\right) \sum_{s, s^{\prime} \in\pm 1} \delta_{n,n'+s}\delta_{m,m'+s}\\ 
            &- P^2 \sum_{s\in\pm 1} \delta_{n,n'+2s}\delta_{m,m'}\left.\right]
        \end{aligned}
    \end{equation}
    \begin{equation}
        \begin{aligned}
            \mathcal{M}_{a_{n,m},b_{n',m'}}=\mathcal{M}_{b_{n,m},a_{n',m'}}=0
        \end{aligned}
    \end{equation}

    \begin{equation}
        \begin{aligned}
            \mathcal{M}_{a_{n,m},x}&=\omega_r\left[-\lambda_0^*\sum_s\phi^0_{n,m+2s}\right.\left.-P\sum_{s,s'}\phi^0_{n+s,m+s'}\right]
        \end{aligned}
    \end{equation}
    \begin{equation}
        \begin{aligned}
            \mathcal{M}_{a_{n,m},y}&=\omega_r\left[-\lambda_0\sum_s\phi^0_{n,m+2s}\right.\left.-P\sum_{s,s'}\phi^0_{n+s,m+s'}\right]
        \end{aligned}
    \end{equation}
    \begin{equation}
        \begin{aligned}
            \mathcal{M}_{b_{n,m},x}&=-\omega_r\left[-\lambda^*_0\sum_s\phi^0_{n,m+2s}\right.\left.-P\sum_{s,s'}\phi^0_{n+s,m+s'}\right]
        \end{aligned}
    \end{equation}
    \begin{equation}
        \begin{aligned}
            \mathcal{M}_{b_{n,m},y}&=-\omega_r\left[-\lambda_0\sum_s\phi^0_{n,m+2s}\right.\left.-P\sum_{s,s'}\phi^0_{n+s,m+s'}\right]
        \end{aligned}
    \end{equation}
    \begin{equation}
        \begin{aligned}
            \mathcal{M}_{x,x}&=(\omega-i\kappa)-E_0\sum_{n,m,s}(\phi^{0}_{n,m})^*\phi^0_{n,m+2s}
        \end{aligned}
    \end{equation}
    \begin{equation}
        \begin{aligned}
            \mathcal{M}_{y,y}&=(-\omega-i\kappa)-E_0\sum_{n,m,s}(\phi^{0}_{n,m})^*\phi^0_{n,m+2s}
        \end{aligned}
    \end{equation}
    \begin{equation}
        \begin{aligned}
            \mathcal{M}_{x,a_{nm}}&=-E_0\lambda_0 \sum_{s}(\phi^0_{n,m+2s})^*-E_0P\sum_{s,s'}(\phi^0x_{n+s,m+s})^*
        \end{aligned}
    \end{equation}
    \begin{equation}
        \begin{aligned}
            \mathcal{M}_{x,b_{nm}}&=-E_0\lambda_0 \sum_{s}\phi^0_{n,m+2s}-E_0P\sum_{s,s'}\phi^0_{n+s,m+s}
        \end{aligned}
    \end{equation}
    \begin{equation}
        \begin{aligned}
            \mathcal{M}_{y,a_{nm}}&=E_0\lambda_0^* \sum_{s}(\phi^0_{n,m+2s})^*+E_0P\sum_{s,s'}(\phi^0_{n+s,m+s})^*
        \end{aligned}
    \end{equation}
    \begin{equation}
        \begin{aligned}
            \mathcal{M}_{y,b_{nm}}&=E_0\lambda_0^* \sum_{s}\phi^0_{n,m+2s}+E_0P\sum_{s,s'}\phi^0_{n+s,m+s}
        \end{aligned}
    \end{equation}
    \begin{equation}
        \begin{aligned}
            \mathcal{M}_{x,y}&=\mathcal{M}_{y,x}=0
        \end{aligned}
    \end{equation}
\clearpage
\bibliography{Bibliography,other}
\end{document}